\documentclass[10pt,conference]{IEEEtran}
\usepackage{cite}
\usepackage{amsmath,amssymb,amsfonts}
\usepackage{graphicx}
\usepackage{textcomp}
\usepackage[usenames,dvipsnames]{xcolor}
\usepackage[hyphens]{url}
\usepackage{fancyhdr}
\usepackage{hyperref}

\usepackage[font=small,labelfont=bf]{caption}
\usepackage{subcaption}
\usepackage{enumitem}
\usepackage{algorithm}
\usepackage{algpseudocode}
\usepackage{listings}
\usepackage{adjustbox}
\usepackage{soul}
\usepackage{multirow}
\usepackage{tikz}
\usepackage{def}
\usepackage[normalem]{ulem}
\usepackage[final]{microtype}
\usepackage{afterpage}
\newlength{\oldtextfloatsep}

\usepackage{comment}
\usepackage{balance}

\newbox\mybox

\usepackage{setspace}

\newcommand{\hpcayear}{2025}

\newcommand{\hpcasubmissionnumber}{112}
\title{\pname{}: Architecting an Efficient Memory-Semantic CXL-based SSD with OS and Hardware Co-design} 

\def\hpcacameraready{} 

\newcommand\hpcaauthors{Haoyang Zhang\footnotemark$^*$, Yuqi Xue\footnotemark$^*$, Yirui Eric Zhou, Shaobo Li, Jian Huang}
\newcommand\hpcaaffiliation{University of Illinois Urbana Champaign}
\newcommand\hpcaemail{\{zhang402, yuqixue2, yiruiz2, shaobol2, jianh\}@illinois.edu}

\fancypagestyle{firstpage}{
  \fancyhf{}
  
  \fancyfoot[C]{\thepage}
}

\author{
  \ifdefined\hpcacameraready
    \IEEEauthorblockN{\hpcaauthors{}}
      \IEEEauthorblockA{
        \hpcaaffiliation{} \\
        \hpcaemail{}
      }
  \else
    \IEEEauthorblockN{\normalsize{HPCA \hpcayear{} Submission
      \textbf{\#\hpcasubmissionnumber{}}} \\
      \IEEEauthorblockA{
        Confidential Draft \\
        Do NOT Distribute!!
      }
    }
  \fi 
}

\fancypagestyle{camerareadyfirstpage}{%
  \fancyhead{}
  
  \fancyhead[C]{
    \ifdefined\aeopen
    \parbox[][12mm][t]{13.5cm}{\hpcayear{} IEEE International Symposium on High-Performance Computer Architecture (HPCA)}    
    \else
      \ifdefined\aereviewed
      \parbox[][12mm][t]{13.5cm}{\hpcayear{} IEEE International Symposium on High-Performance Computer Architecture (HPCA)}
      \else
      \ifdefined\aereproduced
      \parbox[][12mm][t]{13.5cm}{\hpcayear{} IEEE International Symposium on High-Performance Computer Architecture (HPCA)}
      \else
      \parbox[][0mm][t]{13.5cm}{\hpcayear{} IEEE International Symposium on High-Performance Computer Architecture (HPCA)}
    \fi 
    \fi 
    \fi 
    \ifdefined\aeopen 
      \includegraphics[width=12mm,height=12mm]{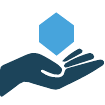}
    \fi 
    \ifdefined\aereviewed
      \includegraphics[width=12mm,height=12mm]{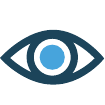}
    \fi 
    \ifdefined\aereproduced
      \includegraphics[width=12mm,height=12mm]{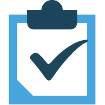}
    \fi
  }
  \fancyfoot[C]{}
}
\begin{document}
\maketitle

\ifdefined\hpcacameraready 
  \thispagestyle{camerareadyfirstpage}
  \pagestyle{empty}
\else
  \thispagestyle{plain}
  \pagestyle{plain}
\fi

\newcommand{\hpcaheight}{0mm}
\ifdefined\eaopen
\renewcommand{\hpcaheight}{12mm}
\fi


\begin{abstract}
The CXL-based solid-state drive (CXL-SSD) provides a promising approach towards scaling the main memory capacity at low cost. 
However, the CXL-based SSD faces performance challenges due to the long flash access latency and unpredictable events such as garbage collection in the SSD device, stalling the host processor and wasting compute cycles. Although the CXL interface enables the byte-granular data access to the SSD, accessing flash chips is still at page granularity due to physical limitations. The mismatch of access granularity causes significant unnecessary I/O traffic to flash chips,
worsening the suboptimal end-to-end data access performance. 

In this paper, we present \pname{}, an efficient CXL-based SSD that employs a holistic approach to address the aforementioned challenges by co-designing the host operating system (OS) and SSD controller. To alleviate the long memory stall when accessing the CXL-SSD, \pname{} revisits the OS context switch mechanism and enables opportunistic context switches upon the detection of long access delays. To accommodate byte-granular data accesses, \pname{} architects the internal DRAM of the SSD controller into a cacheline-level write log and a page-level data cache, and enables data coalescing upon log cleaning to reduce the I/O traffic to flash chips. \pname{} also employs optimization techniques that include adaptive page migration for exploring the performance benefits of fast host memory by promoting hot pages in CXL-SSD to the host.  
We implement \pname{} with a CXL-SSD simulator and evaluate its efficiency with various data-intensive applications. Our experiments show that \pname{} outperforms current CXL-based SSD by 6.11$\times$, and reduces the I/O traffic to flash chips by 23.08$\times$ on average. 
\pname{} also reaches 75\% of the performance of the ideal case that assumes unlimited DRAM capacity in the host, which offers an attractive cost-effective solution. 



\end{abstract}

\renewcommand{\thefootnote}{\fnsymbol{footnote}}
\footnotetext[1]{Co-primary authors.}

\renewcommand{\thefootnote}{\arabic{footnote}}

\section{Introduction}
\label{sec:intro}

CXL-based solid-state drives (CXL-SSDs) have been presented as a practical and cost-effective approach towards expanding the memory capacity~\cite{atc-cxlssd,cxlssd}, as current manufacturing technology has allowed SSDs to scale up to terabytes, and the cost of SSDs is significantly lower than DRAM~\cite{flatflash:asplos2019, flashblox, fleetio:asplos2025, blockflex:osdi2022}. 
The CXL-SSD allows programs to use the SSD as main memory via load/store instructions in a transparent fashion. It enables byte-granular data access to SSDs via CXL protocols. Because of these enabled memory properties in SSDs, we also define CXL-SSD as memory-semantic SSD.  

The CXL technology facilitates the use of flash-based SSDs as memory~\cite{cxlssd,2BSSD, flatflash:asplos2019}. However, simply treating SSDs as an extension of host memory via CXL causes dramatic performance degradation and excessive CPU stalls (see $\S$\ref{sec:motivation}). This is for three major reasons. First, the flash access latency is several orders of magnitude higher than the host DRAM latency. Although the SSD has an internal DRAM cache, the capacity is relatively small (a few GBs in modern SSDs)~\cite{deepstore:micro2019, leaftl:asplos2023}, its miss penalty is still determined by the long flash access latency. Second, due to the inherent properties of flash memory (i.e., out-of-place updates and garbage collection), the complexity of managing flash chips inside the SSD controller causes performance interference. For instance, the garbage collection (GC) of SSDs will postpone the read/write requests to flash chips until the GC is finished. The underlying SSD events can cause long host CPU stalls and unpredictable end-to-end performance. Third, although CXL technology enables byte-granular data access to SSDs, flash chips support only page-granular data access due to physical limitations~\cite{inflash:iscas2018, deepstore:micro2019, 2BSSD}. The mismatch of data access granularity between CXL (byte-granular) and flash chips (page-granular) causes high I/O amplification and extra I/O traffic to flash chips.

To address the above challenges, 
we employ a holistic approach to develop an effective CXL-based SSD, named \pname{}, by co-designing the host OS and CXL-based SSD controller. 
We elaborate the key ideas of \pname{} as follows. 

\noindent
\underline{Coordinated context switch for CXL-SSD.}
To hide memory access latency, processors typically employ out-of-order execution and issue multiple memory requests in parallel in the hope that there are sufficient non-memory instructions to fill the pipeline, while waiting for the response from the memory. This technique has been proven effective with the host DRAM, however, it fails to hide the long flash access latency of CXL-SSDs, unless the processor can examine an impractically large instruction window (for identifying sufficient instructions to hide the memory latency). 
In modern OS, if one thread needs to wait for a long SSD access, the OS can perform a context switch and select another thread to utilize the CPU core. 
Unfortunately, this context switch opportunity is missing for CXL-SSDs, because the OS cannot intercept the load/store memory instructions issued directly from the host CPU to the SSD device via the CXL protocol.

We revisit the OS context switch mechanism and develop a coordinated approach between the host OS and the SSD controller. To precisely track which instruction is blocked by long SSD delay, 
we extend the CXL.mem response packet format to encode a long-delay hint. When the SSD controller detects that a CXL memory request will suffer from a long delay, it responds to the host with this hint. The host CXL controller forwards this hint to the CPU core in the form of a hardware exception triggered by the corresponding load/store instruction. Then, the exception handler calls the host OS scheduler to perform a context switch. \pname{} supports different policies for deciding when to trigger a context switch and which thread will be executed next (see $\S$\ref{sec:design:context_switch}).

\noindent
\underline{CXL-aware SSD DRAM management.} 
Since modern SSDs are primarily designed as block devices, they organize the DRAM cache in page granularity. However, for CXL-SSDs, our study in $\S$\ref{sec:bkg_motiv} finds that most workloads access less than 40\% of cachelines in more than 75\% of pages. Caching the entire page in the SSD DRAM significantly wastes precious SSD DRAM space. 
This also leads to significant write amplification, as we need to write back the entire page to flash chips even though only a few cachelines of a page are dirty.

To bridge the gap between the cacheline-granular CXL interface and the page-granular flash chips, 
we structure the SSD DRAM into a cacheline-granular write log and a page-granular read-write cache. Write requests are served by the write log at cacheline granularity without first fetching the original page from flash chips. The read-write cache is managed in page granularity to exploit spatial locality, as we need to read an entire page from flash chips anyway. When the write log is full, \pname{} performs log compaction in the background to coalesce writes to the same page. This greatly reduces the write traffic to the flash chips and mitigates long flash write latency. For read requests, \pname{} looks up the write log and the read-write cache in parallel to locate the latest data with an efficient hash-based indexing mechanism (see $\S$\ref{sec:design:write_log}).

Since the SSD DRAM capacity is limited, we leverage the host memory to expand the SSD DRAM capacity by enabling adaptive page migrations in the background. 
\pname{} identifies hot pages in the SSD DRAM and performs page migrations transparently (see $\S$\ref{subsec:migration}). 
\pname{} ensures data consistency during page migrations by employing a promotion buffer in the host bridge developed in prior studies~\cite{flatflash:asplos2019}. 
Upon the completion of a page migration, the corresponding page table entry will be updated to reflect the new memory address.

We implement \pname{} with a CXL-SSD simulator based on MacSim~\cite{macsim} and SimpleSSD~\cite{simplessd:micro18}. 
We extend MacSim to simulate context switches on each CPU core and modify its memory interface to simulate the CXL.mem. The CXL memory requests are sent to the SSD that has the write log, the data cache, and the flash translation layer (FTL). 
We evaluate \pname{} with data-intensive workloads (see Table~\ref{tab:benchmarks}). For each workload, we capture the instruction traces of each thread using PIN~\cite{pin} and replay the multi-threaded traces in our simulator. Our evaluation shows that \pname{} 
outperforms state-of-the-art CXL-SSDs by 6.11$\times$, and reduces the I/O write amplification to the flash chips by 23.08$\times$ on average. \pname{} also achieves 75\% of the performance of the ideal case assuming unlimited host DRAM capacity, demonstrating its benefit on cost-effectiveness.
In summary, we make the following contributions:

\begin{itemize}[leftmargin=*]
    \vspace{1ex}
    \item We examine the performance bottlenecks of CXL-SSDs, and identify that they are caused by excessive CPU stalls due to long CXL memory access latency, and the access granularity mismatch between the CXL interface and the flash memory.
    
    \vspace{0.5ex}
    \item We propose \pname{}, which employs an OS and hardware co-design approach to hide the flash access latency of CXL-based SSDs 
    with a coordinated context switch. 
    \vspace{0.5ex}
    \item We re-architect the SSD DRAM cache with a write log and a read-write cache to bridge the gap between the page-granular flash accesses and the byte-granular CXL memory accesses.
    
    \vspace{0.5ex}
    \item We implement \pname{} in a CXL-SSD simulator to accurately simulate the interplay among the CXL-SSD, the multi-core processor microarchitecture, and the OS scheduling.
    
    \vspace{0.5ex}
    \item We evaluate the effectiveness of \pname{} with various data-intensive workloads and sensitivity analysis, showing that \pname{} is a practical and cost-effective approach.
\end{itemize}

\section{Background and Motivation}
\label{sec:bkg_motiv}


\subsection{CXL and Memory Expansion}
\label{sec:cxl_background}

The Compute Express Link (CXL)\mbox{~\cite{cxl-spec}} is a new interconnect standard built on PCIe 5.0 physical interface. It can construct a unified and coherent memory space and enable high-speed communication across different types of processors, memory, and accelerators. CXL has been rapidly gaining industry adoption and is on track to become a primary interconnect. 
CXL defines three protocols that include CXL.io, CXL.cache, and CXL.mem for different purposes. CXL.io is functionally equivalent to the traditional PCIe protocol.
CXL.cache enables cache coherence between interconnected devices. CXL.mem enables the device's local memory to be directly accessed by the host CPU via load/store instructions. With these protocols, CXL supports three primary device types: Type-1 is for devices to access the host memory in a cache-coherent manner (only CXL.cache enabled), such as specialized accelerators like NICs; Type-2 is for devices and the host CPU to access each other's memory with cache coherence (both CXL.cache and CXL.mem enabled);
and Type-3 is for devices that allow the host CPU to access and cache its memory at cacheline granularity (only CXL.mem enabled), such as memory expander devices.

In this paper, we use SSD as a Type-3 CXL device. The entire SSD is exposed as host-managed device memory (HDM).
The SSD memory space is mapped to the host's physical memory space. The CXL.mem protocol allows the host CPU to directly access the SSD via cachable load/store instructions.

\begin{figure}[t]
    \centering
    \includegraphics[width=0.8\linewidth]{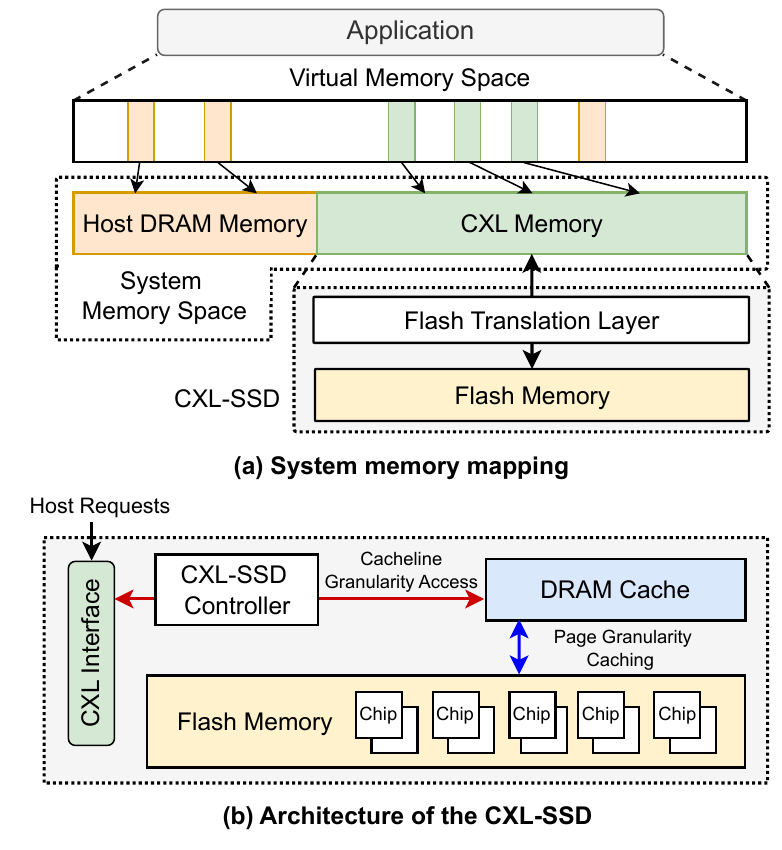}
    \caption{System architecture of CXL-based SSD (CXL-SSD).}
    \vspace{-0.5ex}
    \label{fig:CXLSSD_Type3_Baseline}
\end{figure}

\subsection{Architecture of CXL-based SSDs}
\label{subsec:cxlbaseline}

A simple approach to building a CXL-SSD is to directly connect CXL with a byte-addressable SSD by having a few simple changes to the SSD controller, as described in prior studies~\cite{atc-cxlssd,hello-bytes,cmm-h}. We show its architecture in Figure \ref{fig:CXLSSD_Type3_Baseline}. 

Figure~\ref{fig:CXLSSD_Type3_Baseline} (a) shows the memory mapping with a CXL-SSD. Upon booting, the host initializes the CXL-SSD by mapping its logical memory space into the system physical memory space. The CXL memory or HDM is exposed to the OS as normal physical memory. It is CPU-cacheable and accessible with load/store instructions. However, it possesses different performance attributes compared to the host DRAM. Therefore, the entire system memory can be considered as a heterogeneous memory system. The host OS remains responsible for managing the memory placement and the virtual-to-physical memory mapping. And the Flash Translation Layer (FTL) of the SSD handles the address translation from the logical page address (LPA) to the physical page address (PPA) of flash memory.

Figure~\ref{fig:CXLSSD_Type3_Baseline} (b) shows the architecture of the CXL-SSD. When an application generates a memory access to the CXL-SSD, the CXL home agent will send a message via the CXL.mem protocol. The SSD controller will then parse the message to extract the memory request and serve the data by coordinating with the SSD firmware. To support cacheline (64B) access granularity, the controller utilizes the DRAM cache inside the SSD to serve the memory requests from the host. The SSD DRAM will cache the data from flash chips at page granularity.


\subsection{Challenges of Using CXL-Based SSDs}
\label{sec:motivation}

Although the CXL technology offers a great opportunity for the wide adoption of memory-semantic SSDs,
the current OS and processor architecture does not work well with CXL-SSDs out of the box.
Na\"ively treating CXL-SSDs as conventional DRAM memory will lead to severe performance degradation.

To understand this issue, we study various data-intensive applications (see Table~\ref{tab:benchmarks}) and examine their performance when allocating all their data (1) in a CXL-SSD device as described in $\S$\ref{subsec:cxlbaseline} and (2) in the host DRAM. For each program, we launch four threads on four cores without hyperthreading. We use PIN~\cite{pin} to collect the instruction and memory traces, and replay the traces in a cycle-accurate simulator (see \S\ref{sec:implementation} for details) to quantify the microarchitectural performance impact of using CXL-SSDs. We summarize our key insights as follows.

\begin{figure}[t]
    \centering
    \includegraphics[width=0.85\linewidth]{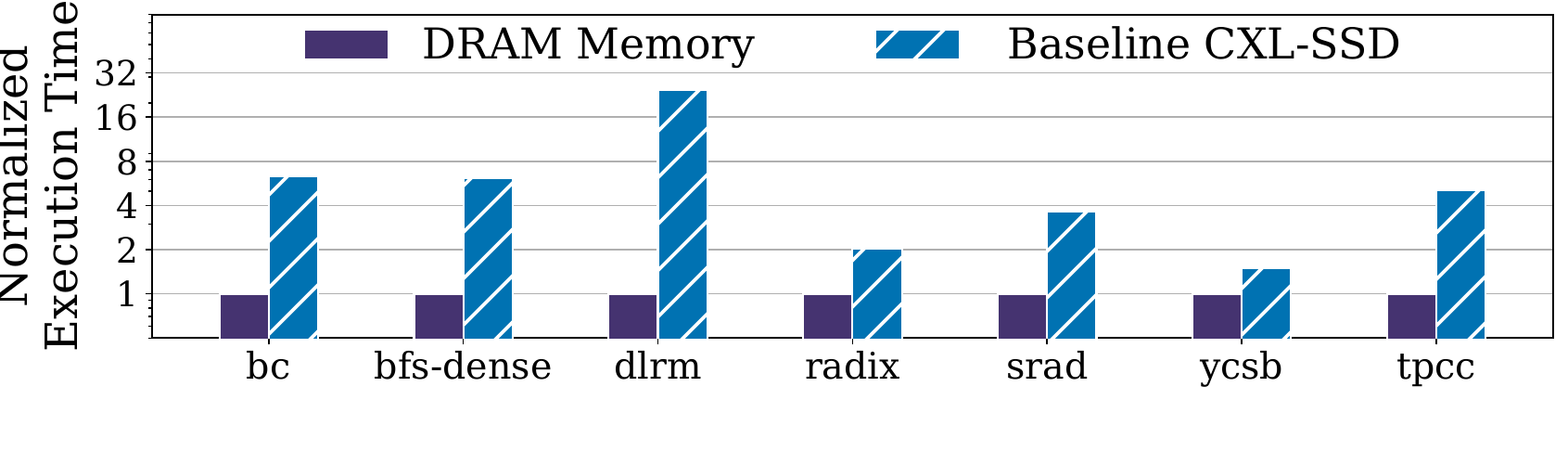}
    \caption{End-to-end execution time of running different workloads using DRAM vs. CXL-SSD.}
    \vspace{-1.5ex}
    \label{fig:motiv_e2e_latency}
\end{figure}


\begin{figure}[t]
    \centering
    \includegraphics[width=0.85\linewidth]{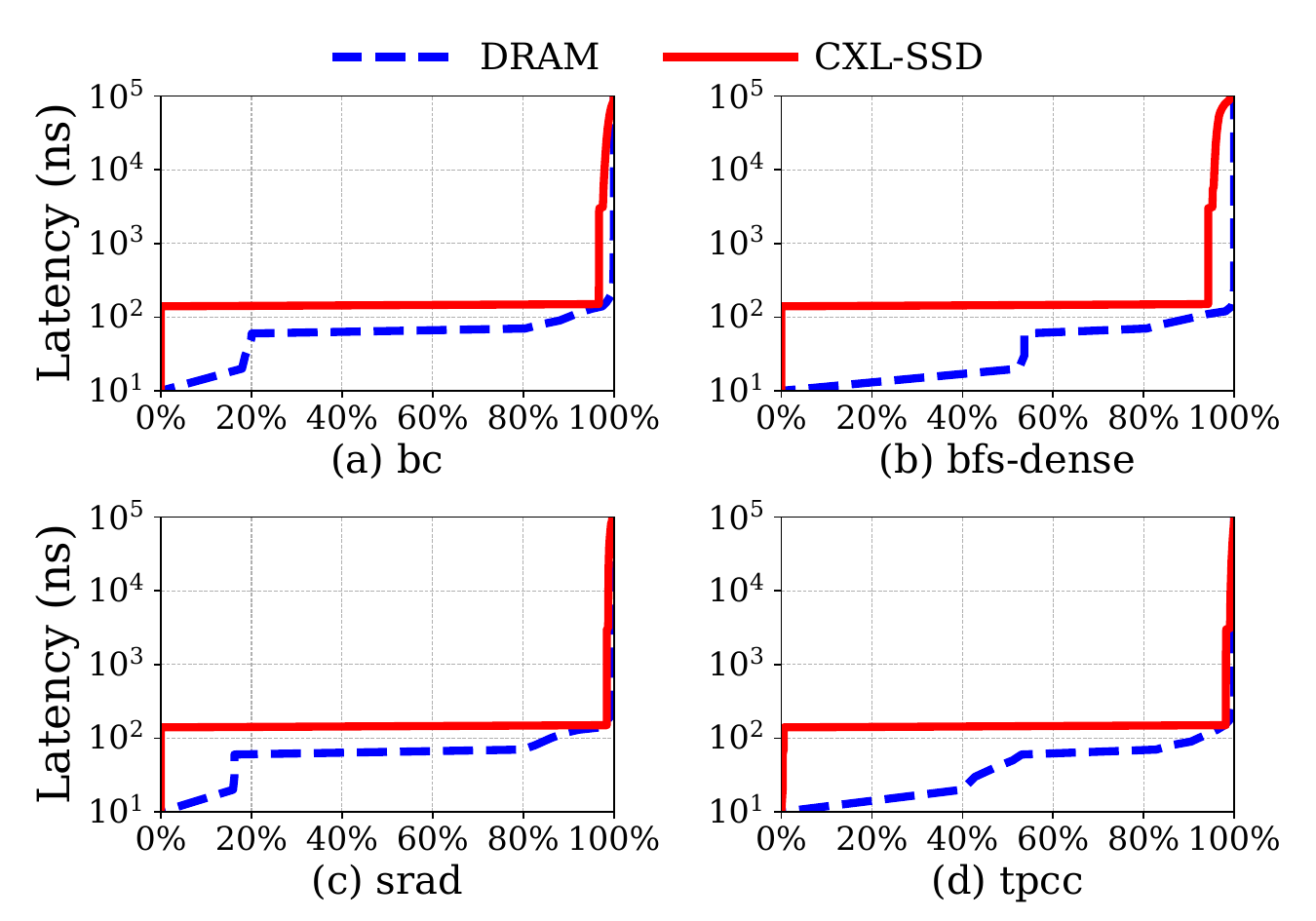}
    \caption{Latency distribution of DRAM vs. CXL-SSD.}
     \vspace{-1.5ex}
    \label{fig:motiv_mem_latency}
\end{figure}

\begin{figure}[t]
    \centering
    \includegraphics[width=0.88\linewidth]{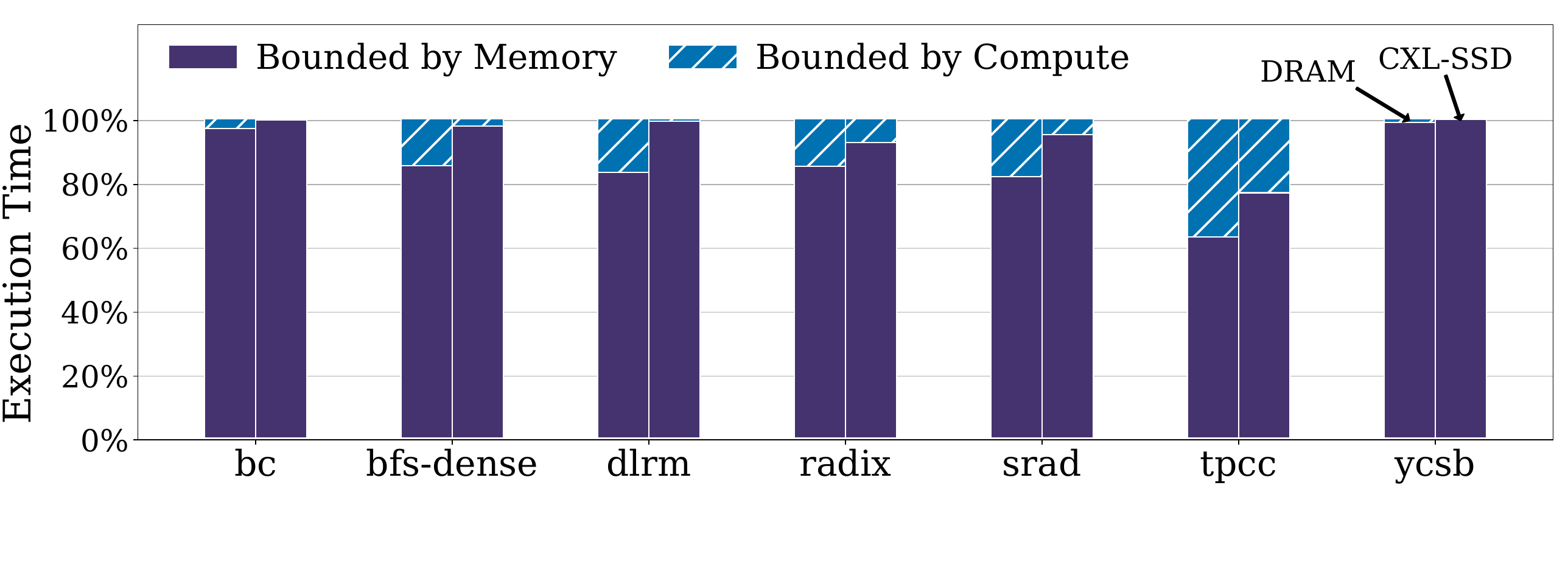}
    \vspace{-1ex}
    \caption{Execution boundedness breakdown of various workloads with DRAM vs. CXL-SSD.}
    \vspace{-1ex}
    \label{fig:motiv_boundedness}
\end{figure}


\noindent
\textbf{Long tail latency.}
Figure~\ref{fig:motiv_e2e_latency} shows the total execution time of the workloads with host DRAM and CXL-SSD, respectively. These workloads perform 1.5--31.4$\times$ worse with CXL-SSD than with DRAM due to the long flash memory access latency even with SSD internal DRAM cache.
Figure~\ref{fig:motiv_mem_latency} shows the off-chip memory access latency distribution of DRAM and CXL-SSD. Due to space limitations, we only show four representative workloads (all workloads have similar patterns).
While more than 90\% of the CXL-SSD memory requests are within 200 ns thanks to the SSD DRAM cache, the tail latency can be as high as hundreds of $\mu$s when a flash read/write happens due to the SSD DRAM cache miss. The latency will be even higher (e.g., a few milliseconds) when garbage collection is triggered. 

\noindent
\textbf{Excessive processor pipeline stalls.}
The tail latency of CXL-SSD causes severe processor pipeline stalls, leading to both performance degradation and underutilization of CPU and SSD bandwidth.
We quantify the impact of pipeline stalls by analyzing compute vs. memory boundedness, following Intel's Vtune profiling tool~\cite{intel_vtune_bound}. We define that a clock cycle is \textit{bounded by memory} if no instructions except memory operations are executing in this cycle (i.e., the pipeline is stalled by memory) and \textit{bounded by compute} otherwise. 
Figure~\ref{fig:motiv_boundedness} quantifies the percentage of cycles bounded by memory or compute.
The portion of memory-bounded cycles grows from 62.9\%--98.7\% with DRAM to 77\%--99.8\% with CXL-SSD. Although modern processors employ techniques such as OoO and multi-level caches to hide memory latency and exploit memory parallelism, they are less effective for hiding the long flash access latency, causing severe pipeline stalls.

Even worse, the long pipeline stalls lead to memory bandwidth underutilization of the CXL-SSD device.
Although we can use more cores to improve the bandwidth, this will also lead to more severe compute underutilization as more cores are being stalled. This is because the processor cannot keep enough in-flight memory requests to saturate the available SSD bandwidth. For example, to saturate a single DDR5 channel with a bandwidth of 32GB/s and 70 ns latency for each 64B cache line, we need to issue at least $70\times 32/64 = 35$ concurrent memory requests. To hide the flash access latency (3 $\mu$s read latency for state-of-the-art Z-NAND~\cite{znand}) assuming a bandwidth of 16 GB/s for PCIe 5.0 x4, we need $3000\times 16/64 = 750$ memory requests, which is impractical for today's processor.





\noindent
\textbf{Access granularity mismatch between CXL interface and flash memory.}
To hide the long flash access latency, modern SSDs typically employ an internal DRAM cache managed at flash page granularity, as they are designed for the block interface, and flash chips support only page-granular access~\cite{inflash:iscas2018, deepstore:micro2019, 2BSSD}. Thus, the current SSD DRAM cache design becomes significantly less effective for the CXL-SSD due to the access granularity mismatch between the CXL interface (64B cache line) and the flash memory (4KB page or even larger).

\begin{figure}[t]
    \centering
    \includegraphics[width=0.82\linewidth]{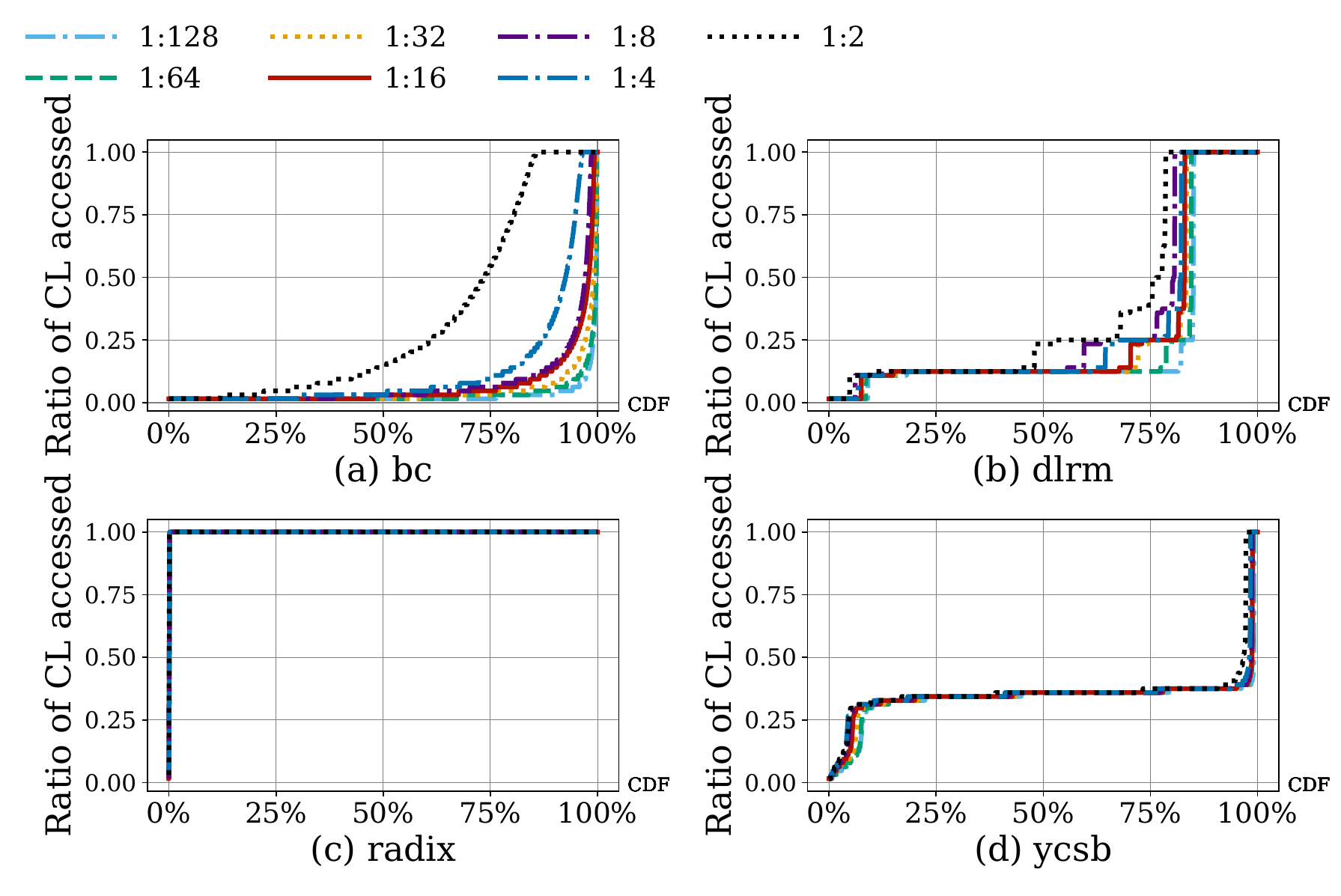}
    \caption{Locality distribution of all pages read from flash chips into the SSD DRAM cache. The legend ``\texttt{1:n}'' means the workload's memory footprint is \texttt{n}$\times$ larger than the SSD DRAM cache. The y-axis is the percentage of cache lines accessed in each page.}
    \vspace{-1ex}
    \label{fig:READ_page_locality}
\end{figure}

\begin{figure}[t]
    \centering
    \includegraphics[width=0.82\linewidth]{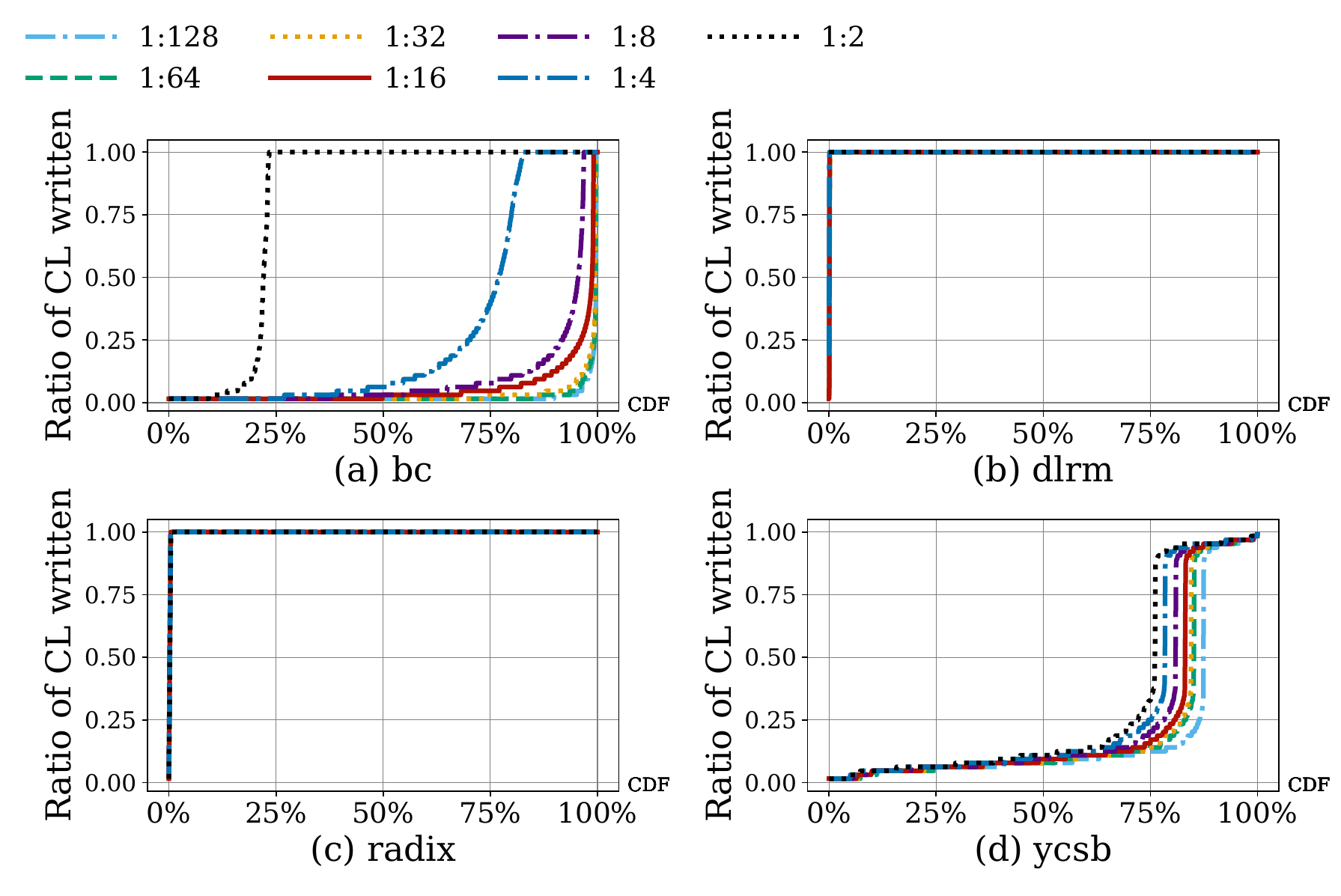}
    \caption{Locality distribution of all pages flushed to flash chips from the SSD DRAM cache. The legend ``\texttt{1:n}'' means the workload's memory footprint is \texttt{n}$\times$ larger than the SSD DRAM cache. The y-axis is the percentage of dirty cache lines in each page.}
    \vspace{-1ex}
    \label{fig:WRITE_page_locality}
\end{figure}

We quantify the memory access patterns of different workloads in Figure~\ref{fig:READ_page_locality} and Figure~\ref{fig:WRITE_page_locality}. Many workloads only access less than 40\% of the cache lines in more than 75\% of pages. This causes two problems. First, as we cache an entire page in the SSD DRAM, the DRAM capacity is significantly wasted as most cache lines in the page are not accessed. Second, even if we only write a few cache lines in a page, we still need to write the entire page to the flash memory, which leads to write amplifications and shortens the SSD lifetime. Enlarging the SSD DRAM capacity has limited benefits unless the DRAM is sufficiently large to hold the entire working set of workloads.


\section{Design of \pname}
\label{sec:design}


\pname{} consists of three major components: (1) the coordinated context switch mechanism based on the detection of long SSD access delays ($\S$\ref{sec:design:context_switch}); (2) a cacheline-granular log-structured memory in SSD controller, for bridging the gap between the byte-granular CXL interface and the page-granular flash chips ($\S$\ref{sec:design:write_log}); (3) an adaptive page migration mechanism that leverages the host memory to expand SSD DRAM by migrating hot pages to the host in a transparent and consistent manner ($\S$\ref{subsec:migration}). We discuss each of them as follows. 

\subsection{Coordinated Context Switch Mechanism}
\label{sec:design:context_switch}

When a thread encounters a long CXL-SSD access caused by SSD DRAM cache miss, we can context switch to another thread to better utilize the CPU core.
However, the SSD device has no knowledge about the microarchitectural status of the host CPU, such as which core triggers a missed memory access and whether this load is speculative.
Similarly, the host CPU does not know whether a memory access is a hit/miss in the SSD DRAM cache.
Neither CPU nor SSD can by itself decide whether to trigger a context switch.
Therefore, to enable context switch on long CXL-SSD memory stalls, we coordinate between host OS and SSD controller. 

To enable a coordinated context switch, we need to decide
(1) when to trigger the context switch; (2) how to conduct the context switch; and (3) what are the policies for the context switch. 
To address these questions, we first present the coordinated context switch procedure. 
Then, we discuss policies for deciding when to trigger a context switch and which thread is executed next. 
Finally, we discuss the hardware and software modifications needed for this mechanism.

\begin{figure}[t]
    \centering
    \includegraphics[width=0.90\linewidth]{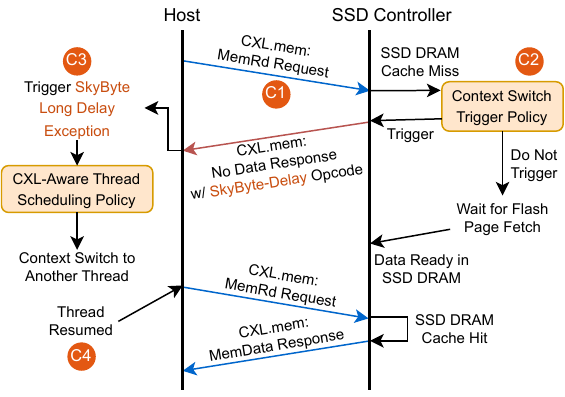}
    \caption{The procedure of coordinated context switch in \mbox{\pname{}}.}
    \vspace{-1ex}
    \label{fig:context_switch}
\end{figure}

\noindent
\textbf{Context switch procedure.}
We show the context switch procedure with an example of a CXL memory read from the host CPU in Figure~\ref{fig:context_switch}. As writes are buffered in the write log (see $\S$\ref{sec:design:write_log}), they do not need to trigger context switch. 

\begin{figure}[t]
    \centering
    \includegraphics[width=0.90\linewidth]{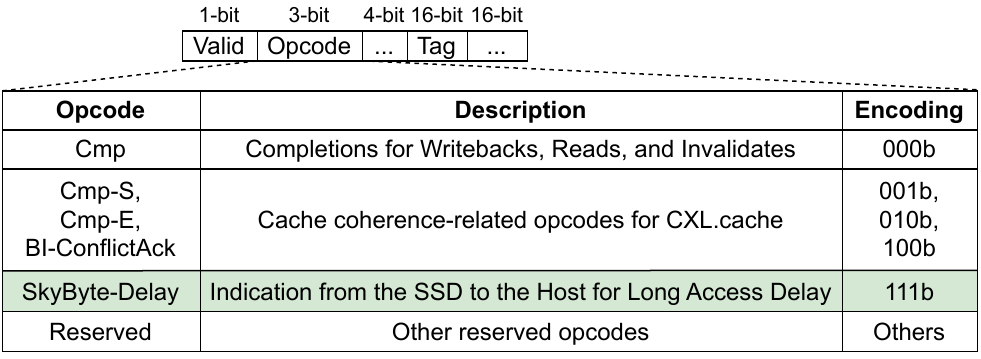}
    \caption{No Data Response (NDR) message format and opcode definitions in CXL.mem. \pname{} uses one of the reserved opcodes (shaded in green) to indicate a long access delay.}
    \vspace{-1ex}
    \label{fig:cxlmem_nodata_response_format}
\end{figure}

\noindent
\circleo{C1} \textul{\textit{Sending CXL.mem request message with tracking information.}}
The host CPU sends a CXL.mem \texttt{MemRd} request message to the SSD controller.
By default, the CPU maintains microarchitectural status, including the miss status handling registers (MSHRs) of the shared LLC, for tracking which load/store instruction in which core is waiting for the response of this memory request.
The MSHRs also perform memory access coalescing, so a memory request may be associated with multiple instructions from different cores if they request for the same cache line. The CXL controller tracks all the memory requests between the host CPU and the SSD via the CXL.mem \texttt{MemRd} message (see Figure~\ref{fig:cxlmem_nodata_response_format}). 


\noindent
\circleo{C2} \textul{\textit{Sending context switch request with extended CXL.mem No Data Response Message.}}
Upon an SSD DRAM cache miss, the SSD controller starts to fetch the page from the flash.
It will determine whether or not to send a context switch request to the host OS based on an estimated access latency (as discussed later in the conetxt switch trigger policy).
The SSD controller sends the context switch request via a No Data Response (NDR) message (one type of the slave-to-master (S2M) message).
The NDR message indicates the completion of a CXL memory request without returning any data to the host CPU.
As shown in Figure~\ref{fig:cxlmem_nodata_response_format}, the \texttt{MemRd} message includes a 16-bit tag~\cite{cxl-spec} for each CXL.mem transaction. \pname{} extends the NDR message specification by introducing a new opcode called \texttt{SkyByte-Delay}.
This opcode indicates that the corresponding \texttt{MemRd} request will suffer from a long access delay (e.g., an SSD DRAM cache miss). 


\noindent
\circleo{C3} \textul{\textit{Triggering context switch with hardware exception.}}
\pname{} leverages the existing hardware exception mechanism in modern CPUs to precisely track which load/store instruction in which core should trigger a context switch. \pname{} defines a new \textit{SkyByte Long Delay Exception}.
Once the host CPU receives the \texttt{SkyByte-Delay} NDR message from the CXL controller, it looks up the LLC MSHR entry of this memory request and traverses the upper-level cache hierarchy (e.g., L1 and L2) to find all uncommitted memory instructions waiting for this response. 
When any of these instructions enter the retire stage, it will trigger the SkyByte Long Delay Exception (similar to how a load/store instruction triggers a Page Fault Exception) on the corresponding core. The address of this instruction will be saved upon the context switch, such that when the thread is switched back, it will resume from this instruction and re-issue this memory access to the CXL-SSD. 

Such a design eliminates false-positive context switches, where a load/store triggers a context switch but is later squashed, at no extra hardware cost, since modern processors by default delay the exception handling to the retire stage. For example, speculative load/store and hardware prefetch will not trigger any exception even if they miss in the SSD DRAM.

\mbox{\pname{}} installs a special handler for the SkyByte Long Delay Exception in the x86 interrupt descriptor table (IDT). The exception handler invokes a CXL-aware thread scheduling policy to decide which thread is executed next and performs the context switch, which will be discussed later. 


\noindent
\circleo{C4} \textul{\textit{Resuming the original thread.}}
When the original thread is scheduled back, it will resume from the previously missed memory instruction that triggered the \mbox{\pname{}} Long Delay Exception. This memory access is then issued again to the CXL-SSD. If the replayed instruction hits in the SSD DRAM, the data will be returned with a CXL.mem \texttt{MemData} response.





When a thread is context-switched, all its pending load/store instructions are squashed. 
However, the
MSHRs in the cache hierarchy may or may not be freed depending on the implementation~\cite{inform_mem_op:isca96, nemirovsky2022multithreading}.
This can cause severe MSHR contention between threads given the long flash access latency.
In the worst case, a thread performs accesses to 64 cache lines in multiple 4KB pages, all of which miss in the SSD DRAM cache. These requests will occupy at least 64 MSHRs for a few microseconds, easily exhausting all the MSHRs. To address this issue, we free the MSHR entry as soon as the corresponding instruction squashes. Since this approach can also benefit host DRAM accesses, we enable it in \pname{} by default.


When a thread is scheduled again after it has been context-switched away a while ago, it may trigger the same SSD DRAM cache miss again if the requested page has already been evicted due to a cache conflict.
This is less of a concern, since the LRU eviction policy in the SSD DRAM cache already prevents the requested page from being evicted for most of the time. For instance, in our experiments with various data-intensive applications, we did not observe a page being evicted before the original thread resumes execution and accesses it. 
\noindent
\textbf{Context switch trigger policy.}
Upon an SSD DRAM cache miss, the SSD controller can choose to either trigger a context switch or let the host CPU wait for the data. Intuitively, if the context switch overhead is smaller than the CXL-SSD access delay, we can perform a context switch to hide the delay.

\pname{} uses a \textit{threshold-based policy} to decide whether a context switch should be triggered.
The SSD controller estimates the latency of fetching the requested page from flash chips. If the estimated latency is higher than the threshold, a context switch will be triggered. The threshold can be tuned based on the host context switch overhead. 

We show the details of the threshold-based context switch trigger policy in Algorithm~\ref{algo:threshold_trigger_policy}. 
\pname{} first looks up the FTL mapping table to get the physical page address (PPA), which determines which flash channel this request will be issued to (Line 2--3). It employs a simple approach to estimate the flash access latency by querying the corresponding flash controller's queue status, i.e., the number of requests placed in the queue (Line 4). Typically, the requests in the channel queue will be served in FIFO order\mbox{\cite{ssd-book}}. Therefore, similar to the approach studied in prior work~\cite{mqsim, femu}, \pname{} can accurately estimate the delay of the request by summing the latency of all requests in the queue (Line 5--6). 
If the estimated delay is longer than the threshold, \pname{} will trigger a context switch (Line 7). If a request is blocked by an ongoing garbage collection (GC), \mbox{\pname} will immediately trigger a context switch, as GCs typically last for milliseconds. While the GC process will block the issuing of requests in the queue, 
its impact is already considered in the latency prediction algorithm by querying the flash channel queue status.

\setlength{\textfloatsep}{0ex}
\begin{algorithm}[t]
 \footnotesize
\caption{Threshold-based context switch trigger policy.}
\label{algo:threshold_trigger_policy}
\begin{algorithmic}[1]
\Function{shd\_ctx\_swtc} {\textit{req}, \textit{threshold}, \textit{read\_lat}, \textit{write\_lat}, \textit{erase\_lat}}
    \State{\textit{PPA} = translate\_address(\textit{req});}
    \State{\textit{queue} = get\_channel\_queue(\textit{PPA});}
    \State{\textit{num\_read}, \textit{num\_write}, \textit{num\_erase} = \textit{queue}.get\_counters();}
    \State{\textit{est\_lat} = \textit{read\_lat} * (\textit{num\_read} + 1) + \textit{write\_lat} * \textit{num\_write}}
    \State{\;\;\;\;\;\;\;\;+ \textit{erase\_lat} * \textit{num\_erase};}
    \State{\textbf{return} \textit{est\_lat} $>$ \textit{threshold};}
\EndFunction
\end{algorithmic}
\end{algorithm}
\afterpage{\global\setlength{\textfloatsep}{\oldtextfloatsep}}

\begin{figure}[t]
    \centering
    \includegraphics[width=0.82\linewidth]{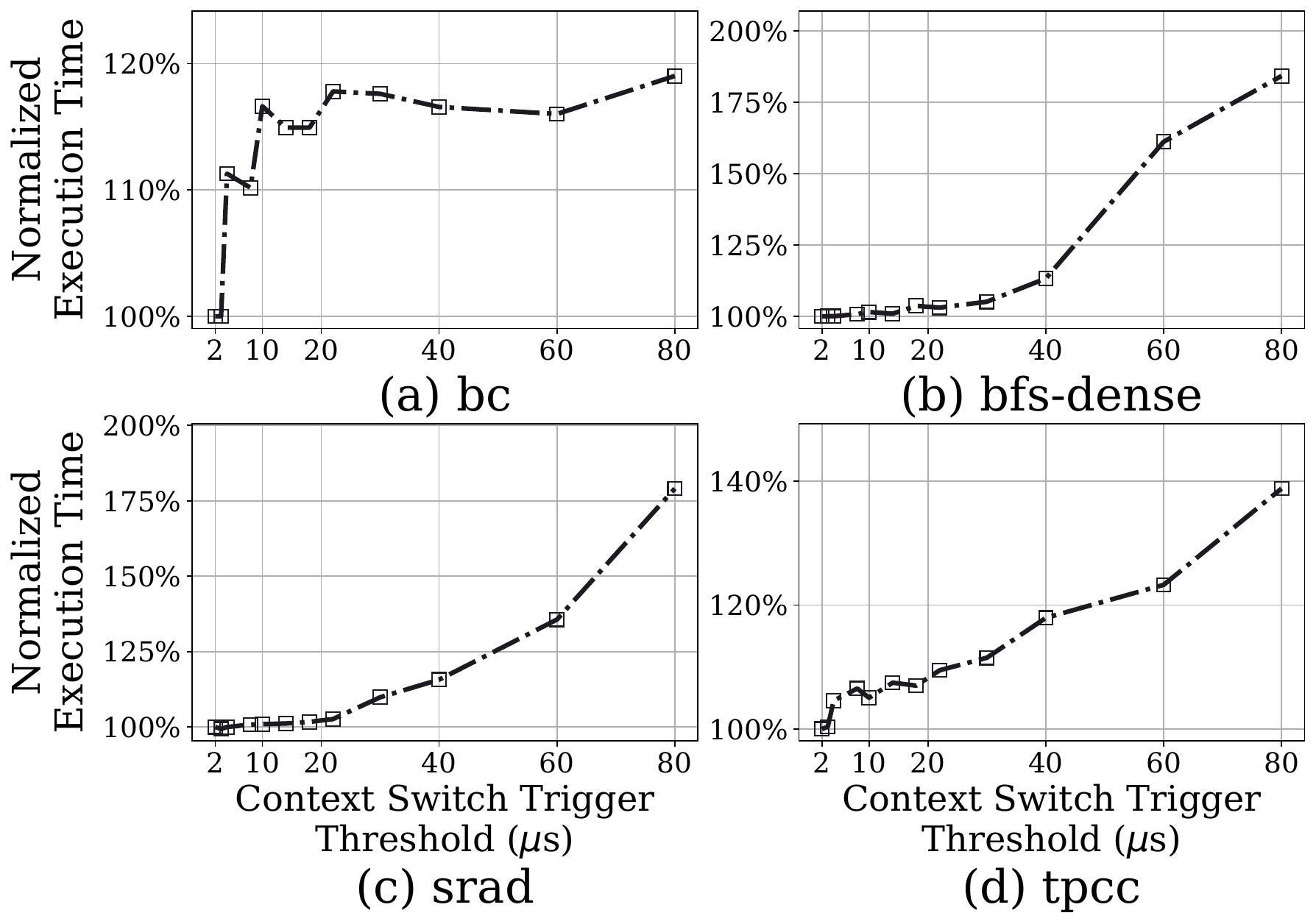}
    \caption{Impact of the thresholds defined in the in the coordinated context switch trigger policy.}
    \label{fig:context_switch_trigger_threshold}
\end{figure}

To set an appropriate threshold, we can measure the average overhead of context switches of the host CPU. 
Figure~\ref{fig:context_switch_trigger_threshold} shows the performance of various thresholds for representative workloads. Since the flash page read latency (3 $\mu$s by default) of our SSD is longer than the regular context switch overhead (2 $\mu$s, as examined with the hardware setup described in Table~\ref{tab:exp_setup}), we set the threshold to 2 $\mu$s. In practice, the threshold can be tuned empirically for different CPUs and system configurations. \pname{} allows the host OS to configure it.





\noindent
\textbf{OS support for CXL-aware context switch.} 
To enable the Long Delay Exception, we install a new exception handler into the x86 IDT. When the exception is raised, the exception handler yields the CPU resources owned by the current thread. The system scheduler then decides the thread to run next based on the predefined policy. The yield thread is re-enqueued back to the run queue in OS, allowing it to be scheduled again later.



We explore three scheduling policies for the scheduler to pick the next runnable thread in Linux OS:
\textit{(1) Round-Robin (RR) policy}, in which
the threads take turns to execute;
\textit{(2) Random policy}, in which 
a thread is chosen randomly to execute next;
\textit{(3) Complete Fairness Scheduler (CFS) policy}~\cite{cfs_schedular}, in which 
it prioritizes the thread which has the shortest received execution time for ensuring a fair share of CPU cycles among threads. 

\begin{figure}[t]
    \centering
    \includegraphics[width=0.90\linewidth]{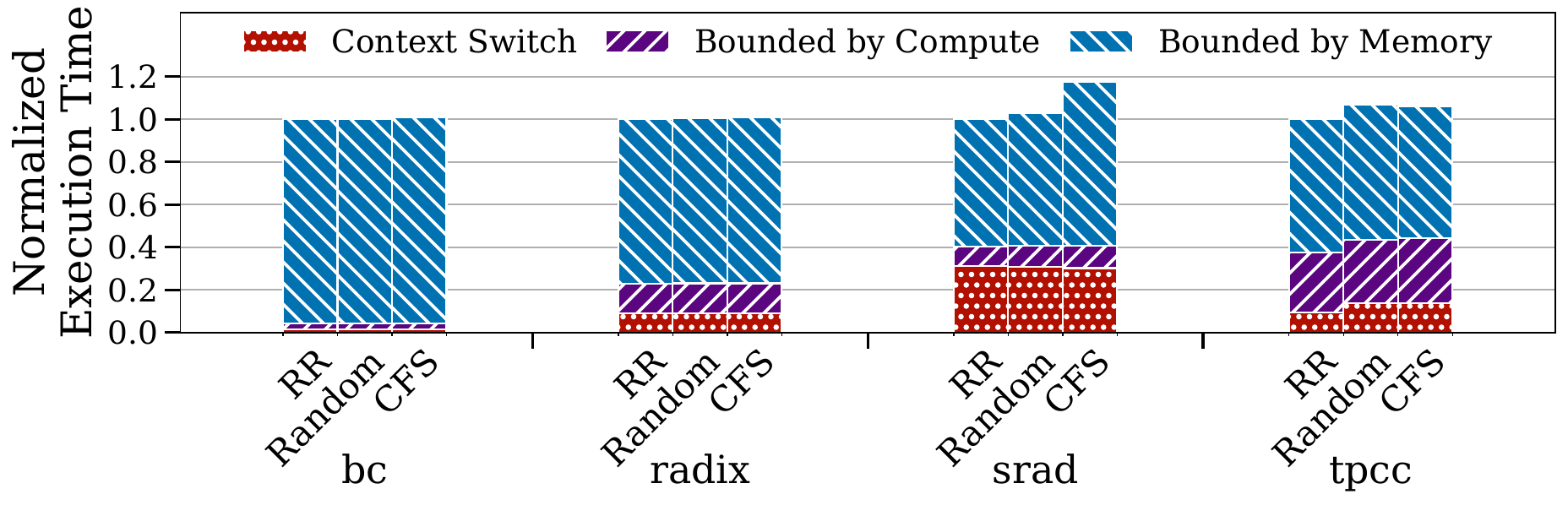}
    \caption{Normalized execution time of \mbox{\pname{}} with different thread scheduling policies (lower is better). }
    \vspace{-2ex}
    \label{fig:eval_context_switch_policy}
\end{figure}

We evaluate the performance of these policies and show the execution time breakdown in Figure \mbox{\ref{fig:eval_context_switch_policy}}, following the same definition of boundedness in Figure \mbox{\ref{fig:motiv_boundedness}}. 
The three policies deliver similar performance. This is because all the threads are mainly bounded by memory I/O, and all three policies allow the threads to have equal or similar opportunities to issue memory requests to the SSD, even though these threads may trigger a context switch immediately after they are scheduled. This improves the SSD bandwidth utilization with multi-threading (see \mbox{$\S$\ref{sec:eval_context_switch}}). For some workloads (e.g., \texttt{srad}), context switching takes a considerable amount of time because context switches are repeatedly triggered by all threads when they are all waiting for flash accesses.  The CFS policy may perform slightly worse in a few workloads, because it needs to enforce fair sharing for all the threads, which may cause the OS to select the threads that have just been scheduled away due to the context switch in \pname{}. But its impact on the end-to-end performance is trivial. Since CFS has become a standard scheduling policy in modern OSes like Linux, we employ it by default in \mbox{\pname{}}.
%

\subsection{CXL-Aware SSD DRAM Management}
\label{sec:design:write_log}

As discussed in \mbox{$\S$\ref{sec:motivation}},  the internal DRAM cache of SSDs can be suboptimal due to the mismatch of the access granularity between the CXL interface and the flash memory. 
A cacheline request would trigger a flash page access, which will cause a long delay, and waste precious SSD DRAM capacity. 

\begin{figure}[t]
   \centering
   \includegraphics[width=0.83\linewidth]{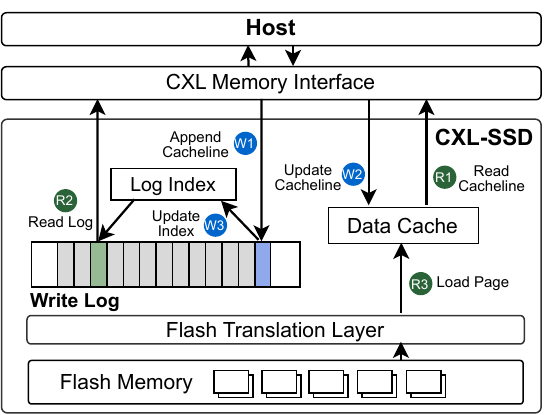}
   \caption{The architecture of CXL-aware SSD DRAM.}
   \label{fig:cxl-ssd-structure}
\end{figure}

To address these issues, we re-architect the SSD DRAM, as shown in Figure~\ref{fig:cxl-ssd-structure}.
\pname{} deploys a cache-line granular \textit{double-buffered write log} to buffer all write requests from the host. All cacheline writes are directly appended to the log without flash access along the critical path. They are flushed to the flash memory later. Compared to a page-granular cache, the write log has two benefits: (1) caching at a finer granularity saves the precious SSD DRAM space when the write locality is bad, and (2) the write log provides a larger write coalescing window, which reduces the flash write traffic. We maintain a \textit{log indexing table} to index the latest data in the write log array for read requests or log compaction.

For read operations, when flash access is inevitable upon an SSD DRAM miss, we still need to fetch the entire page. Thus, \pname{} caches the fetched page with a read-write data cache managed in page granular to exploit data locality. To maintain data consistency between the write log and data cache, we update both the write log and cache upon a write request, and during a read, we check the write log first. 
To speed up the requests, \pname{} performs parallel updates and parallel lookups on both sides.
The write log and the data cache use logical addresses for indexing, as they are built on top of the Flash Translation Layer (FTL). We discuss them as follows. 

\begin{figure}[t]
    \centering
    \includegraphics[width=0.83\linewidth]{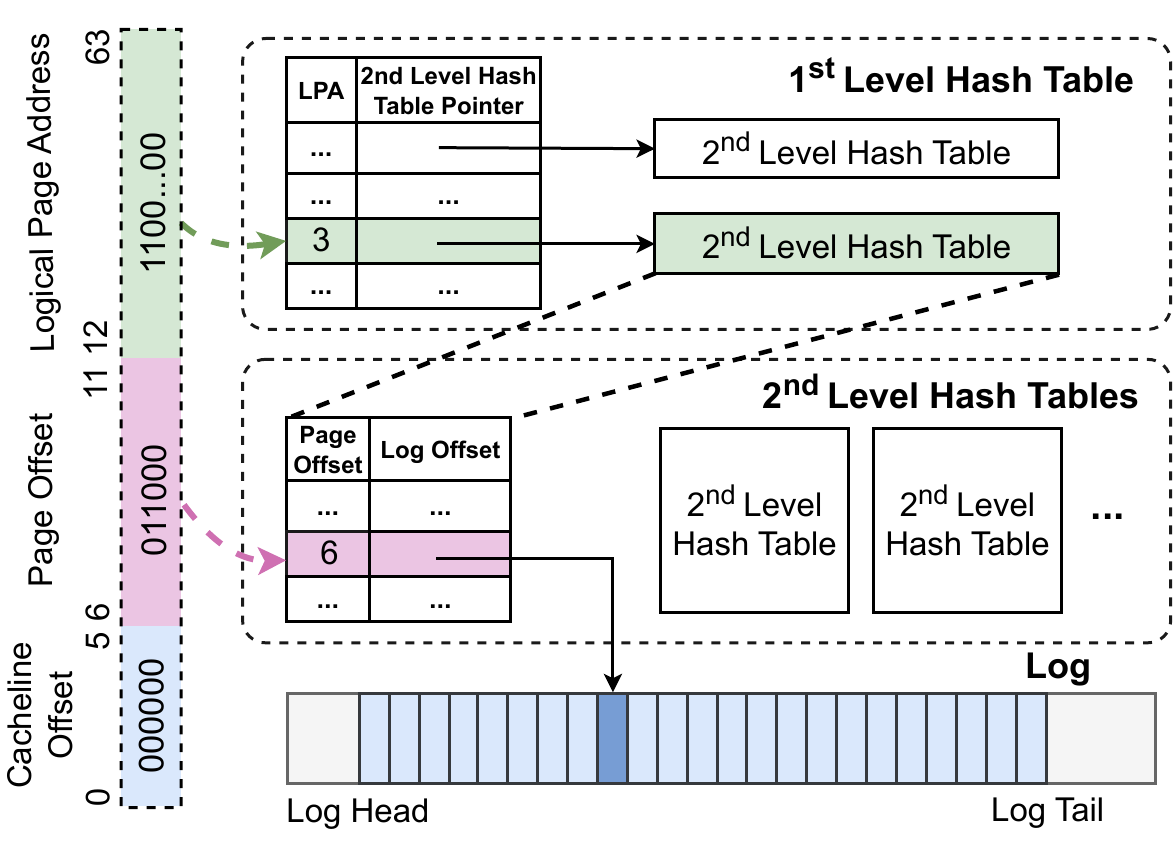}
    \caption{The two-level hash table structure for indexing the write log.}
    \vspace{-1ex}
    \label{fig:indexing_of_log}
\end{figure}

\noindent
\textbf{Write log structure.}
Figure~\ref{fig:indexing_of_log} shows the structure of the write log. The log records all written 64B cache lines within a circular buffer with head and tail pointers.
To achieve a fast lookup of log entries, \pname{} uses the hash table for indexing, as it provides an amortized \texttt{O(1)} lookup latency. 
However, a plain hash table will randomly distribute the stored entry, and it would require multiple lookups when finding all cache lines within the same page during compaction. To solve the problem, \pname{} breaks the indexing structure into two levels.
At the first level, we employ a hash table indexed by the logical page address (LPA). Each valid entry points to a \textit{second level hash table} that tracks all logged cache lines in this logical flash page with their offset within the page. The log offset of the tracked cache line in the log is indexed in the second-level table. We can easily find all updated cache lines in the same page by traversing the corresponding second-level hash table.

In the first-level hash table, each entry stores the 8B LPA and 8B second-level hash table pointer for each page. Since each page (4KB) has 64 cache lines, we only need 6 bits to index the page offset. The second-level hash table entry is 4B, which includes the 6-bit page offset and a 26-bit log offset. 
Consider a 64MB write log with 1M entries, if all second-level hash tables are initially allocated with full 64 entries, it requires 272MB of memory under the worst case when each page only contains a single dirty cache line. To reduce the memory footprint, we instead allocate small-sized second-level hash tables and allow them to resize on demand. \pname{} initiates each second-level hash table with four entries (16B), and doubles the table size whenever the table load factor exceeds a threshold (0.75 by default). With resizing, under the worst case, our memory footprint only occupies up to 32MB (1M 16B first-level hash entries and 1M 16B second-level hash tables).
Our experiments with real workload traces ({$\S$\ref{sec:eval}}) show that the log index occupies 5.6MB of memory on average. 

\noindent
\textbf{Read operation.}
We show the read/write operation in Figure~\ref{fig:cxl-ssd-structure}. When a read request arrives, the SSD controller looks up both the data cache and the write log in parallel. If the requested data is cached, we directly read from the data cache and return to the host (\circler{R1}). When the data cache misses but the write log holds the cache line, \pname{} will retrieve it from the write log (\circler{R2}). 
When both write log and data cache miss, the entire page is fetched from the flash into the data cache, and \pname{} returns the target cache line (\circler{R3}). The write log may contain the recently updated cache lines, and we need to keep the cached page up-to-date. After fetching the page to the cache, \pname{} performs a lookup on the first-level hash table. If it contains the entry of this page, we traverse its second-level table to merge all cache lines in the log to the fetched page.

\noindent
\textbf{Write operation.}
On a write request, \pname{} directly appends the written cache line at the tail of the write log (\circlemw{W1}). An update to the data cache is issued in parallel if it contains the corresponding page (\circlemw{W2}). \pname{} will also update the indexing table (\circlemw{W3}). If the write updates an existing cache line in the write log, \mbox{\pname{}} updates the index table entry pointing to the newest log offset.
\pname{} serves multiple read/write requests in parallel and leverages lock-free hash tables~\cite{lock_free_hash} and queues for synchronization.

\begin{figure}[t]
    \centering
    \includegraphics[width=0.83\linewidth]{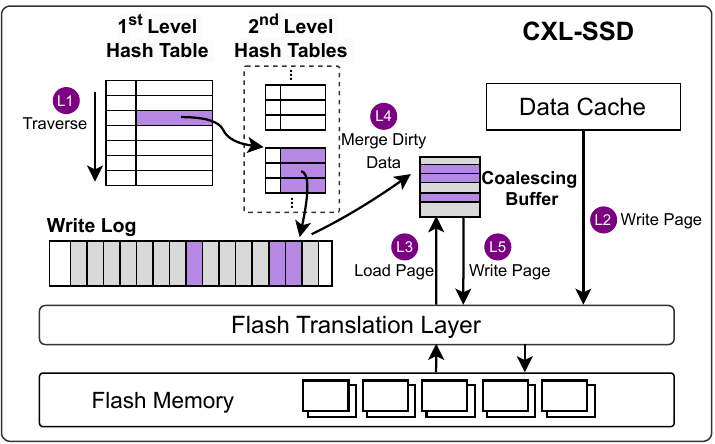}
    \caption{Log compaction in CXL-SSD.}
    \vspace{-2ex}
    \label{fig:log_compaction}
\end{figure}

\noindent
\textbf{Write log compaction.}
Figure~\ref{fig:log_compaction} shows how \pname{} performs the log compaction. To reduce write traffic to the flash chips, \mbox{\pname{}} coalesces the writes during log compaction. Since we only track the newest data in the indexing table, the old updates will be dropped during the compaction.
\pname{} scans the first level hash table to find all pages that need to be flushed (\circlec{L1}). For each page, if it is cached, we directly flush the cached page back to flash memory (\circlec{L2}). Otherwise, we load the missing page from the flash memory to a coalescing buffer (\circlec{L3}). \pname{} then traverse the second level table entries for dirty cache lines in this page and merge them with the loaded page (\circlec{L4}). The merged page is then written back to the flash (\circlec{L5}). When performing flash writes, \mbox{\pname{}} batches the pages in the write buffer, and distributes writes across multiple channels to exploit SSD channel parallelism.

\pname{} maintains a double-buffered log to avoid blocking incoming requests and performs the compaction in the background. When one log becomes full, \mbox{\pname{}} triggers the log compaction process and switches to a new log. During compaction, incoming write requests will be directed to the new log, following the normal write procedure. An incoming read request requires a parallel lookup in both the new log and the old log for the latest data. As the background compaction is not on the critical path of serving memory requests, it does not introduce much overhead, and a single compaction takes 146 $\mu$s on average.
After compaction, we remove the indexing table and reclaim the memory used by the previous log. 

\subsection{Adaptive Page Migration}
\label{subsec:migration}



As the SSD DRAM size is limited, we use the host memory to expand SSD DRAM cache. \pname{} develops an adaptive page migration mechanism to migrate frequently accessed pages to the host. To decide which pages to migrate, \pname{} uses a similar policy developed in prior work~\cite{flatflash:asplos2019, linuxkernel, dong, thremostat, PagePlacement, tap, os_ccnuma}. 
The SSD controller tracks the access count of flash pages and selects pages whose access counts exceed a threshold as the migration candidates.
\pname{} only migrate pages in the SSD DRAM cache, as it includes the candidate hot pages. 
After choosing the target page for migration, the SSD triggers the migration by sending a PCIe MSI-X interrupt with the SSD page address to the host. The host OS then allocates a physical page in the host DRAM memory with the default buddy allocator and copies the page content to the new page.

We need to ensure data consistency during page migrations. To achieve this, a simple approach is to leverage the OS techniques by setting the page under migration as not presented in PTE before the migration, and revising the page fault handler to resume the request after the migration. However, this causes high performance overheads.  
In \pname{}, we follow the prior approach~\cite{flatflash:asplos2019} and track the migration progress with a Promotion Look-aside Buffer (PLB) in the root complex. The PLB has 64 entries, each entry (24B) records all ongoing migrations with the addresses of source/destination pages (8B each), a bitmap for the migrated cache line (8B), and a valid bit. 
Therefore, a read request to a page under promotion can be served from the SSD DRAM. 
For writes, if the migrated bit has been set for the requested cache line, the request is forwarded to the most recent copy of that cache line in the host DRAM.

After data migration, the host OS modifies the page table entry (PTE) to map the original virtual address to the new host DRAM page, with the corresponding TLB entry also updated.
The host OS will acknowledge the migration request, and the SSD removes the page from the data cache and invalidates the write log index by setting the corresponding entry as \texttt{NULL}.   



Since the host DRAM space is limited, \pname{} also enables the host to evict pages back to SSD for free space. \mbox{\pname{}} leverages the existing page reclamation \mbox{policy~\cite{linux_page_reclaim}} in Linux to select the page for eviction, finding a relatively ``cold" page tracked by the active/inactive \mbox{list~\cite{reclaim_list}}.
We then allocate a new page in the CXL memory space and perform the page copy. The host OS will update the corresponding page table entry to point at the SSD page and update the TLB.
\section{Discussion}
\label{sec:discussion}

\noindent
\textbf{Data persistence support.} By default, \mbox{\pname{}} assumes the user application does not require data persistence, so it can transparently promote pages to the host DRAM for performance improvement.
To support data persistence, \mbox{\pname{}} offers the option for users to pin memory pages in the CXL-SSD, such that these pages will not be promoted to the host.
Programmers can use \texttt{clwb} instruction to ensure a cacheline has reached the battery-backed SSD DRAM.

\noindent
\textbf{Support for NUMA architecture.} \mbox{\pname} can work with a multi-socket NUMA machine.
The CXL-SSD appears to the system as a ``CPU-less" NUMA node.
All processors treat the CXL-SSD as non-local memory. The CXL-SSD is attached to the PCIe slot of one CPU socket, referred to as the ``home node''.
Accesses from other NUMA nodes to the CXL-SSD may experience slightly higher latency compared to accesses from the home node, but since the inter-socket latency is much smaller (less than 100 ns) than the flash latency ($\mu s$-level), \mbox{\pname{}} uses the same context switch threshold for all NUMA nodes.
Since the SSD controller is unaware of which NUMA nodes have accessed each page, when a page migration is triggered, the page is migrated to the home node first. If the home node has no free memory, the page will be migrated to the NUMA node with the most free memory space.
After that, if the page is needed by other nodes, the NUMA balancing mechanism of the OS is responsible for further migrations.

\noindent
\textbf{Support for multiple page sizes.} The adaptive page migration of \pname{} can support multiple page sizes (e.g., 2MB huge pages).
When the host OS receives the MSI-X interrupt from the SSD with a 4KB-page address, it first checks whether this address belongs to a huge page. If it does, the host OS allocates a physical huge page in the host DRAM, and migrates the entire page by copying all 4KB data chunks from the SSD. Once the migration finishes, the host sends a custom NVMe command to notify the SSD to remove all corresponding 4KB chunks from its internal DRAM caches.

To migrate a huge page, the PLB needs to track all cachelines in the huge page.
However, tracking all 32,768 cachelines in a 2MB page requires a 4KB bitmap per PLB entry.
To reduce the hardware cost of PLB, we extend the original PLB into a two-level structure.
The first-level entry contains a 64B bitmap that indicates whether a 4KB chunk in a 2MB page has been migrated. The second-level entry contains an 8B bitmap to track which cachelines in a 4KB chunk have been migrated.
The PLB migrates the huge page chunk-by-chunk, and it only needs one first-level entry to track the 2MB page and one second-level entry to track the current 4KB chunk under migration.
\section{Implementation}
\label{sec:implementation}

\noindent
\textbf{\pname simulation framework.}
We implement \pname{} with a cycle-accurate simulator based on MacSim~\cite{macsim} and SimpleSSD~\cite{simplessd:micro18}. 
MacSim replays multi-threaded instruction traces captured by Intel's PIN tool~\cite{pin} on multiple simulated CPU cores.
To implement \pname{}'s coordinated context switch mechanism, we extend it to support task scheduling by selecting which set of instructions to execute on the simulated cores based on the scheduling policy discussed in \S\ref{sec:design:context_switch}. 
We measured the context switch overhead with an Intel E5 CPU under real systems, and we set our experiment timing model accordingly (see Table~{\ref{tab:exp_setup}}). We also simulate the side effects of context switching, including cache contention or branch mispredictions. 
We implement the PLB (\mbox{$\S$\ref{subsec:migration}}) in MacSim and enable the CPU to issue CXL memory requests to the SSD. 
To simulate the side effects of page migration, we also modified MacSim to perform a TLB shootdown for all cores when a page finishes migration.
To simulate the CXL.mem interface and flash accesses in SSDs, we extend MacSim's memory controller logic to redirect all CXL.mem requests to the SSD simulator. The SSD simulator simulates the SSD firmware, including the core FTL functions (e.g., address translation and GC), \pname{}'s write log and data cache in the SSD DRAM, and flash accesses. 

\noindent
\textbf{FPGA SoC prototype.}
We prototype the write log and SSD DRAM cache (\mbox{$\S$\ref{sec:design:write_log}}) on a Xilinx Zynq UltraScale+ ZU3EG MPSoC board.
The board features a quad-core ARM Cortex A53 processor commonly deployed in SSD controllers\mbox{\cite{armStoragesolutions}}, an LPDDR4 memory, and programmable logic resources.
We implement critical-path operations such as indexing the write log and data cache on FPGA. Other off-critical-path tasks like log compaction and garbage collection are managed by the ARM cores.
We use a red-black tree to index the page-granular data cache.
The average lookup latency is 72 ns for a 64 MB write log and 49 ns for a 512 MB data cache. 
The latency grows slightly (by less than 10 ns) with a larger log size.
Our prototype achieves a peak throughput of 11.93/9.37 GB/s for cacheline reads/writes.
We verified the performance model used in our simulator with these measurements.
\section{Evaluation}
\label{sec:eval}

\begin{table}[t]
\scriptsize
\caption{Benchmarks used in our experiments.}
\centering
\begin{tabular}{cccccc}
\cline{1-6}
\multicolumn{1}{|c|}{\multirow{2}{*}{Category}}  & \multicolumn{1}{c|}{\multirow{2}{*}{Suite}}  & \multicolumn{1}{c|}{\multirow{2}{*}{Name}} & \multicolumn{1}{c|}{Memory} & \multicolumn{1}{c|}{Write}  & \multicolumn{1}{c|}{LLC} \\
\multicolumn{1}{|c|}{} & \multicolumn{1}{c|}{} & \multicolumn{1}{c|}{} & \multicolumn{1}{c|}{Footprint} & \multicolumn{1}{c|}{Ratio} & \multicolumn{1}{c|}{MPKI} \\ \cline{1-6}
\multicolumn{1}{|c|}{Graph}  & \multicolumn{1}{c|}{Rodinia~\cite{rodinia}}  & \multicolumn{1}{c|}{\texttt{bfs-dense}} & \multicolumn{1}{c|}{9.13GB} & \multicolumn{1}{c|}{25\%} & \multicolumn{1}{c|}{122.9}\\ 
\multicolumn{1}{|c|}{Processing}  & \multicolumn{1}{c|}{GAP~\cite{beamer2017gap}}  & \multicolumn{1}{c|}{\texttt{bc}} & \multicolumn{1}{c|}{8.18GB} & \multicolumn{1}{c|}{11\%} & \multicolumn{1}{c|}{39.4}\\ \cline{1-6}
\multicolumn{1}{|c|}{HPC}  & \multicolumn{1}{c|}{Splashv3~\cite{splashv3}}  & \multicolumn{1}{c|}{\texttt{radix}} & \multicolumn{1}{c|}{9.60GB} & \multicolumn{1}{c|}{29\%} & \multicolumn{1}{c|}{7.1} \\ \cline{1-6}
\multicolumn{1}{|c|}{Image}  & \multicolumn{1}{c|}{\multirow{2}{*}{Rodinia~\cite{rodinia}}}  & \multicolumn{1}{c|}{\multirow{2}{*}{\texttt{srad}}} & \multicolumn{1}{c|}{\multirow{2}{*}{8.16GB}} & \multicolumn{1}{c|}{\multirow{2}{*}{24\%}}  & \multicolumn{1}{c|}{\multirow{2}{*}{7.5}} \\ 
\multicolumn{1}{|c|}{Processing}  & \multicolumn{1}{c|}{}  & \multicolumn{1}{c|}{} & \multicolumn{1}{c|}{} & \multicolumn{1}{c|}{} & \multicolumn{1}{c|}{} \\ \cline{1-6}
\multicolumn{1}{|c|}{\multirow{2}{*}{Database}}  & \multicolumn{1}{c|}{\multirow{2}{*}{WHISPER~\cite{whisper1.0}}}  & \multicolumn{1}{c|}{\texttt{ycsb}} & \multicolumn{1}{c|}{9.61GB} & \multicolumn{1}{c|}{5.0\%} & \multicolumn{1}{c|}{92.2}\\
\multicolumn{1}{|c|}{}  & \multicolumn{1}{c|}{}  & \multicolumn{1}{c|}{\texttt{tpcc}} & \multicolumn{1}{c|}{15.77GB} & \multicolumn{1}{c|}{36\%}  & \multicolumn{1}{c|}{1.0} \\ \cline{1-6}
\multicolumn{1}{|c|}{Machine}  & \multicolumn{1}{c|}{\multirow{2}{*}{DLRM~\cite{DLRM19}}}  & \multicolumn{1}{c|}{\multirow{2}{*}{\texttt{dlrm}}} & \multicolumn{1}{c|}{\multirow{2}{*}{12.35GB}} & \multicolumn{1}{c|}{\multirow{2}{*}{32\%}}  & \multicolumn{1}{c|}{\multirow{2}{*}{5.1}} \\
\multicolumn{1}{|c|}{Learning}  & \multicolumn{1}{c|}{}  & \multicolumn{1}{c|}{\texttt{}} & \multicolumn{1}{c|}{} & \multicolumn{1}{c|}{}  & \multicolumn{1}{c|}{} \\ \cline{1-6}
\end{tabular}
\label{tab:benchmarks}
\end{table}


Our evaluation shows that:
(1) \pname{} outperforms state-of-the-art CXL-SSD designs by 6.11$\times$ on average ($\S$\ref{sec:eval_overall});
(2) \pname{} scales the performance with more threads by exploiting the context-switching opportunities to hide SSD access latency ($\S$\ref{sec:eval_context_switch});
(3) \pname{} reduces the average memory access time by 14.19$\times$ and the flash write traffic by 23.08$\times$ over the state-of-the-art CXL-SSD design ($\S$\ref{sec:eval_ssd_cache});
(4) \pname{}'s write log design uses the SSD DRAM capacity efficiently ($\S$\ref{sec:eval_write_log});
(5)\pname{} benefits CXL-SSDs with various internal DRAM sizes and flash chip latencies ($\S$\ref{sec:eval_ssd_dram_size} and $\S$\ref{sec:eval_ssd_lat}). 

\subsection{Experiment Setup}
\label{sec:exp_setup}

\noindent
\textbf{Workloads.}
We evaluate representative multi-threaded data-intensive workloads of various domains, including scientific computing, machine learning training, and database queries, as shown in Table\mbox{~\ref{tab:benchmarks}}.
All benchmarks are configured to have a total memory footprint of at least 8GB.
The write ratio of the workloads ranges from 5\% to 36\%, and the LLC misses per kilo-instructions (MPKI) range from 1.0 to 122.9.
We run the workloads on a server with Intel Xeon processors and capture the instruction traces for each thread using Intel's PIN tool~\cite{pin}.
The traces for \texttt{ycsb} and \texttt{tpcc} are collected using the in-memory database \texttt{nstore} \cite{nstore}, and we use workload B in \texttt{ycsb}. 
For each workload, we capture the  traces for at least 100 million instructions per thread, and replay them in our simulator.
In our experiments, all data are initially stored in CXL-SSD.
We use the traces to warm up the simulator, including the CPU caches, the host memory, the SSD DRAM cache, and the write log.
Page tables and program binaries are in the host memory. We precondition the SSD to ensure garbage collections will be triggered.

\begin{table}[t]
\scriptsize
\caption{Parameters defined in our CXL-SSD simulator.}
\centering
\begin{tabular}{|c|c|c|}
    \hline
    \multirow{5}{*}{CPU} & Cores & 8 cores, 4.0 GHz, 256 ROB entries per core \\\cline{2-3}
    & L1 I/D Cache & 32/32 KB, 8/8 ways per core, 8 MSHRs \\\cline{2-3}
    & L2 Cache & 512 KB, 32 ways per core, 128 MSHRs \\\cline{2-3}
    & L3 Cache & 16 MB, 16 ways, shared by all cores, 1024 MSHRs \\\cline{2-3}
    & \multirow{2}{*}{DRAM} & DDR5 4800 MHz, 8 channels, 36-38-38 \\
    & & Max. Total Size of Promoted Pages: 2 GB \\\hline
    \multirow{8}{*}{SSD} & Interface & CXL over PCIe 5.0 x4 (16 GB/s, 40 ns protocol latency) \\\cline{2-3}
    & \multirow{3}{*}{Organization} & 16 channels, 8 chips/channel, 8 dies/chip, \\
    & & 1 plane/die, 128 blocks/plane, \\
    & & 256 pages/block, 4KB page (Total: 128 GB) \\\cline{2-3}
    & \multirow{2}{*}{Flash Latency} & Read ($t_\text{R}$): 3 $\mu$s, Program ($t_{\text{Prog}}$): 100 $\mu$s, \\ 
    & & Erase ($t_{\text{BERS}}$): 1000 $\mu$s \\\cline{2-3}
    & \multirow{2}{*}{DRAM} &  LPDDR4 3200 MHz, 2 channels, 16-18-18 \\ 
    & & Data Cache Size: 512 MB, 2048 MSHRs\\\cline{2-3}
    & GC Policy & Threshold: 80\%, \# of Blocks to Erase: 19660 \\\hline
    \multicolumn{3}{|c|}{Context Switch Overhead: 2 $\mu$s; Context Switch Trigger Threshold: 2 $\mu$s} \\
        \multicolumn{3}{|c|}{Write Log: 64 MB; Data Cache: 448 MB} \\\hline
\end{tabular}
\label{tab:exp_setup}
\vspace{-1.5ex}
\end{table}

\noindent
\textbf{System configurations.}
Table~\ref{tab:exp_setup} lists the simulator parameters.
Following Samsung's recent 2TB CXL-SSD prototype with 16GB DRAM~\cite{cxlssd-16gb, Samsung-fms22}, we scale down the configuration with the same flash-to-DRAM capacity ratio (i.e., 128GB flash with 512MB DRAM cache, excluding the FTL mapping table cache), as it is impractical to simulate a TB-scale SSD at cache line granularity. We use the NAND flash latency of Samsung Z-SSD SZ985, which uses ultra-low-latency (ULL) flash chips. Since modern servers usually have 64GB or larger DRAM (4$\times$ the DRAM size of Samsung's CXL-SSD), we configure the maximum host DRAM size for storing migrated pages as 2GB (4$\times$ our simulated SSD DRAM size). We also conduct sensitivity analysis on SSD DRAM cache size and flash latency.


We conduct an ablation study by testing \pname{} with subsets of design components enabled.
When the coordinated context switch mechanism is enabled, we run 24 threads on 8 cores for all workloads.
We also conduct sensitivity analysis by varying the number of threads.
For other cases, we run 8 threads on 8 cores since more threads will not improve the performance.
We run each workload multiple times with different numbers of threads. In each run, we collect a separate trace for each thread. We ensure that all traces represent the same section of the program.
We compare the following designs:
\begin{itemize}[leftmargin=*]
    \item \textbf{\basename{}}: the state-of-the-art CXL-based SSD device that incorporates all optimizations developed in recent works\mbox{~\cite{atc-cxlssd,hello-bytes}}, including prefetching from flash to SSD DRAM, optimized cache replacement policy, and MSHRs in the SSD controller for tracking accesses to the flash chips.
    \item \textbf{\mbox{\pname{}-C}}: \mbox{\basename{}} with \mbox{\pname{}'s} coordinated context switch mechanism.
    \item \textbf{\pname{}-P}: \basename{} with adaptive page migration.
    \item \textbf{\pname{}-W}: \basename{} with \pname{}'s CXL-aware SSD DRAM management (the write log and data cache).
    \item \textbf{\mbox{\pname{}}-CP}: \mbox{\basename{}} with both \mbox{\pname{}'s} coordinated context switch mechanism and adaptive page migration.
    
    \item \textbf{\pname{}-WP}: \basename{} with both page migration and CXL-aware SSD DRAM management.
    \item \textbf{\pname{}-Full}: the complete version of \pname{} that enables context switches upon \pname{}-WP.
    \item \textbf{DRAM-Only}: the ideal case assuming we run the workload with infinite host DRAM.
\end{itemize}

\subsection{Overall Performance Improvement}
\label{sec:eval_overall}

\begin{figure}[t]
    \centering
    \includegraphics[width=0.92\linewidth]{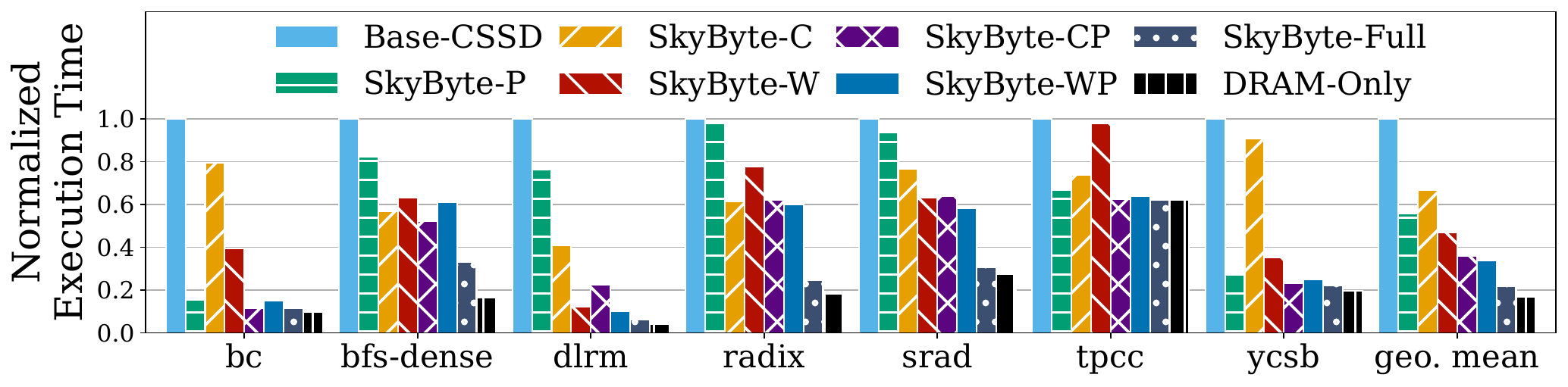}
    \caption{Normalized execution time of \mbox{\pname{}} variants over \mbox{\basename{}} (lower is better).}
    \label{fig:eval_e2e_perf}
\end{figure}

Figure~\ref{fig:eval_e2e_perf} shows the performance of different \pname{} variants normalized to \basename{}.
Overall, \pname{}-Full outperforms \basename{} by up to 16.35$\times$ (6.11$\times$ on average).

\pname{}-P's performance reflects the benefit from page promotions ($\S$\ref{subsec:migration}). It outperforms \basename{} by 1.84$\times$ on average by using host memory to expand the SSD DRAM cache size.
Workloads with better locality (e.g., \texttt{bc}, \texttt{tpcc}, and \texttt{ycsb}) benefit more from page promotions, while other workloads (e.g., \texttt{radix} and \texttt{srad}) benefit less. 

\pname{}-W's performance reflects the benefit of the write log ($\S$\ref{sec:design:write_log}). It outperforms \basename{} by 2.16$\times$ on average.
By buffering and coalescing write requests, \pname{}-W significantly reduces the writeback traffic to the flash chips and hence reduces the overall latency of flash access requests.
The benefit is especially obvious for temporally sparse workloads like \texttt{bc} and \texttt{dlrm}, which are otherwise frequently bottlenecked by write misses in \basename{}.
Workloads that have many sparse writes (e.g., \texttt{srad}) benefit more from \pname{}-W, while workloads that are sensitive to the DRAM cache size and have relatively fewer sparse writes (e.g., \texttt{tpcc}) benefit more from \pname{}-P.
When combining both optimizations, \pname{}-WP outperforms \basename{} by 2.95$\times$ on average.


\begin{figure}[t]
    \centering
    \includegraphics[width=0.95\linewidth]{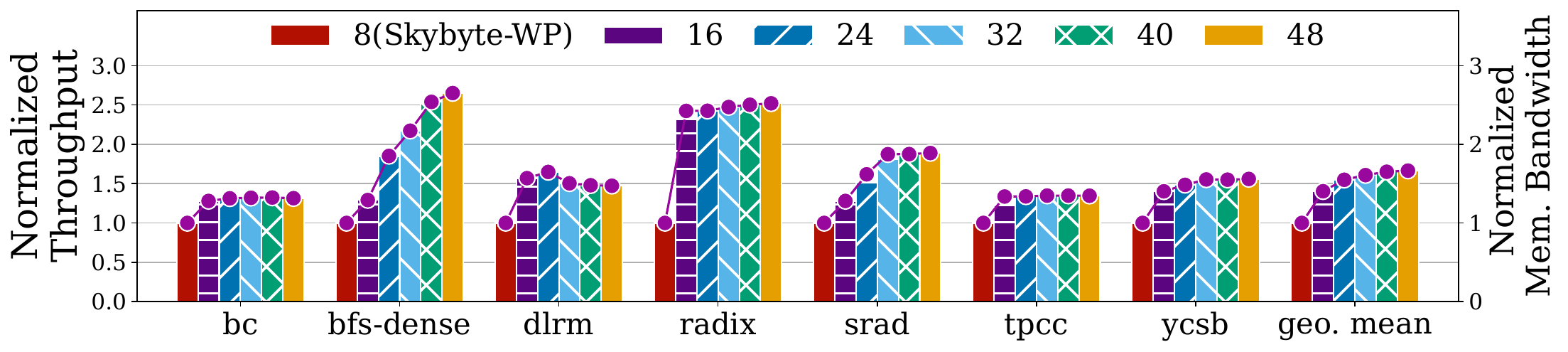}
    \caption{Throughput (bars) and SSD bandwidth utilization (lines) of \pname{}-Full as we increase the number of threads (normalized to \pname{}-WP with 8 threads).}
    \label{fig:eval_thpt_vary_threads}
      \vspace{-1.5ex}
\end{figure}


\mbox{\pname{}-CP} leverages host DRAM as a cache and uses context switches to hide I/O latency. It outperforms \mbox{\basename{}} by 2.79$\times$. \mbox{\pname{}-Full} further outperforms \mbox{\pname{}-CP} by 1.64$\times$ on average, as it also utilizes the write log to address the granularity mismatch issue of CXL-SSD.

\pname{}-Full outperforms \pname{}-WP by 1.55$\times$ on average and \basename{} by 1.61--16.35$\times$.
While \pname{}-WP already represents a well-optimized CXL-SSD device, its ideal performance is still bottlenecked by the long flash accesses.
\pname{}-Full enables opportunistic context switches and leverages multi-threading to further hide long flash delays.
By comparing the performance improvement of \pname{}-C over \basename{} and that of \pname{}-Full over \pname{}-WP (both are 1.49$\times$ on average), we observe that the benefit of context switching is orthogonal to that of write log and page migration, except for \texttt{tpcc}, where the benefit of page migration dominates.

Compared to infinite host DRAM (\texttt{DRAM-Only}), \pname{}-Full achieves 75\% of the ideal performance with CXL-SSD. Given the unit cost of DDR5 DRAM~\cite{DRAM_price} and ULL SSD~\cite{pm1735} is \$4.28/GB and \$0.27/GB respectively based on their market price in summer 2024, \pname{}-Full costs 15.9$\times$ less than the DRAM-only setup and improves cost-effectiveness by 11.8$\times$.

\subsection{Benefit of Context Switch with Varying Number of Threads}
\label{sec:eval_context_switch}



\begin{figure}[t]
    \centering
    \includegraphics[width=0.98\linewidth]{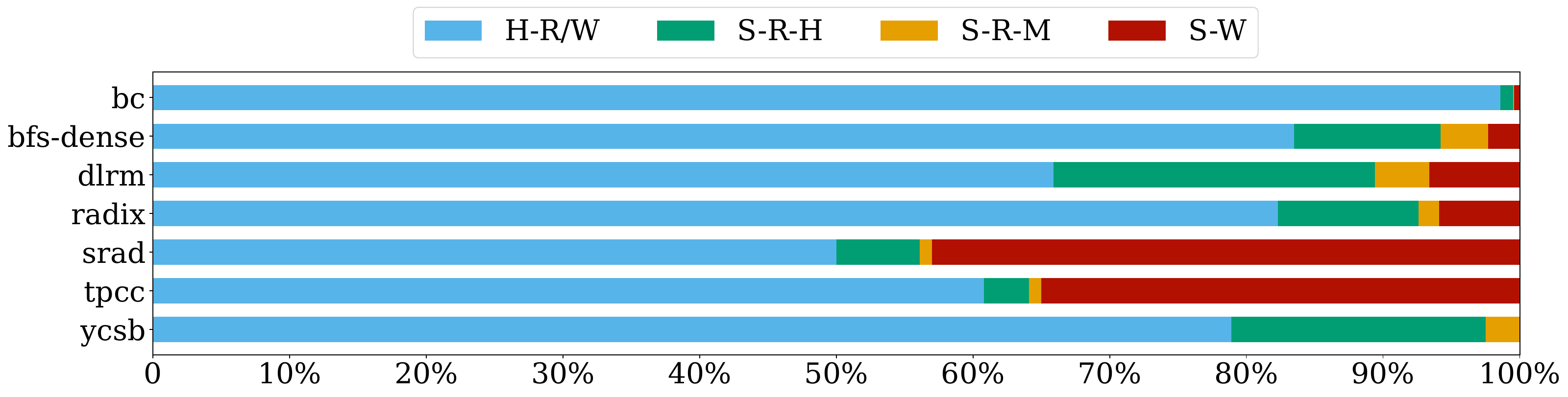}
	\caption[Caption for LOF]{Breakdown of all memory requests of \mbox{\pname{}\footnotemark[1]}. H-R/W: host DRAM read/write. S-R-H: CXL-SSD DRAM read hit. S-R-M: CXL-SSD DRAM read miss. S-W: CXL-SSD write.}
    \label{fig:breakdown_access}
\end{figure}

We analyze the throughput improvements of the coordinated context switch mechanism ($\S$\ref{sec:design:context_switch}) by varying the number of threads with \pname{}-Full on 8 CPU cores, as shown in Figure~\ref{fig:eval_thpt_vary_threads}.
The throughput is strongly correlated to the memory bandwidth utilization improvement with multi-threading. The potential of such improvement depends on the flash read latency (Table~\ref{tab:wp_read-latency}) and the percentage of flash reads over all memory accesses (Figure~\ref{fig:breakdown_access}). If most accesses are absorbed by the host and SSD DRAM, context switching among multiple threads has limited benefits. Otherwise, with more flash accesses, the throughput scales linearly with the number of threads because of higher SSD bandwidth utilization. The benefit diminishes once the context switch overhead exceeds the flash read latency.

Table~\ref{tab:wp_read-latency} and Figure~\ref{fig:breakdown_access} shows the average latency and the percentage of flash reads for \pname{}-WP.
Workloads that suffer from more flash accesses and higher average flash read latency (e.g., \texttt{srad}, \texttt{bfs-dense}) benefit more from more threads.
If the workload frequently triggers log compaction, the long flash writes will interfere with the flash reads, hence increasing the read latency.
In this case, more threads can better hide the long latency. On the other hand, if the average flash read latency is already close to the context switch latency (e.g., for \texttt{bc} and \texttt{dlrm} in Table~\ref{tab:wp_read-latency}), having two threads per core is sufficient to achieve optimal throughput. 
For some workloads (e.g., \texttt{dlrm}), the throughput drops when the context switch overhead becomes high with too many threads.
With more threads, the latency of accessing flash chips may increase due to queuing delays. 
However, the end-to-end performance is improved, since the SSD bandwidth utilization is higher.

\begin{table}[t]
\footnotesize
\caption{Average flash read latency of \pname{}-WP.}
\centering
\begin{tabular}{|c|c|c|c|c|c|c|c|}
\hline
Workload & bc   & bfs-dense & dlrm & radix & srad  & tpcc  & ycsb \\ \hline
Latency ($\mu s$) & 3.5 & 25.7     & 3.4 & 4.9  & 22.5 & 19.6 & 3.3 \\ \hline
\end{tabular}
\label{tab:wp_read-latency}
\end{table}

\footnotetext[1]{As we focus on the memory requests served by the CXL-SSD, we do not distinguish host reads and writes. We also do not distinguish CXL-SSD write hits and misses, as all CXL-SSD writes will append to the write log.}

\begin{figure}[t]
    \centering
    \begin{subfigure}{\linewidth}
        \centering
        \includegraphics[width=0.95\linewidth]{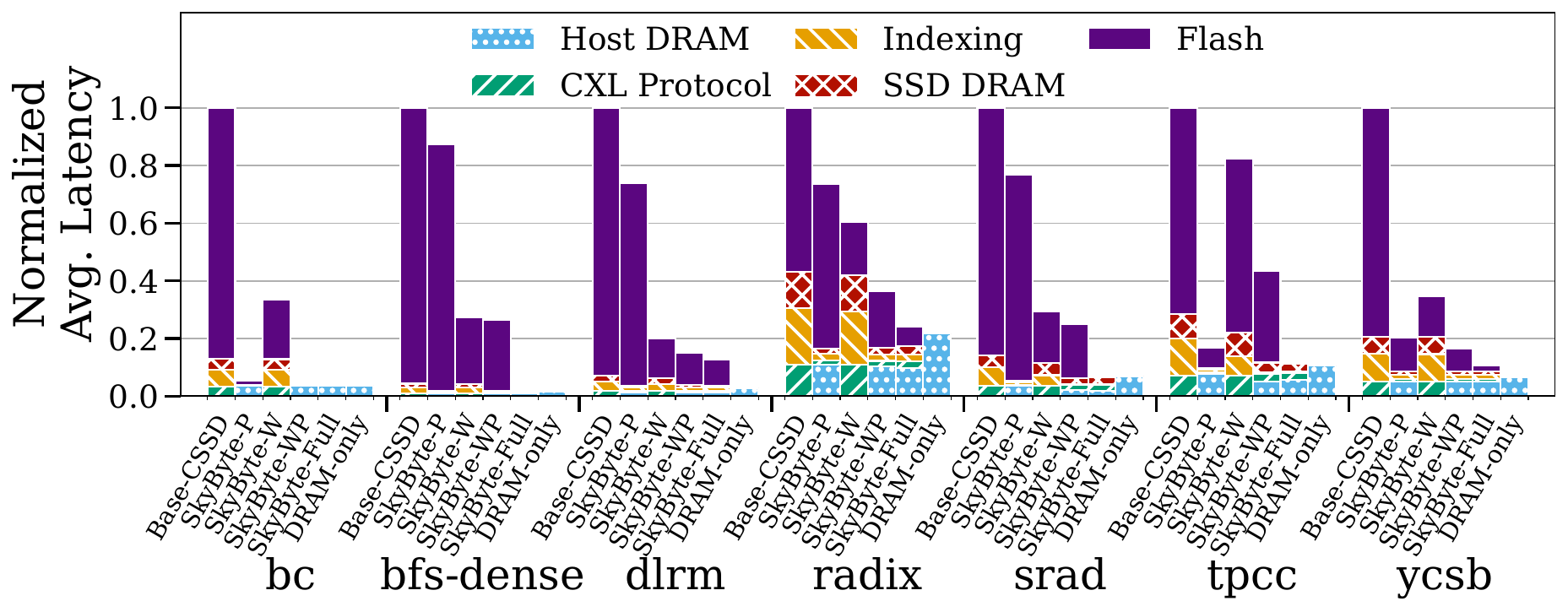}
        \caption{Average memory access time (normalized to \basename{}).}
    \end{subfigure}
    \vspace{0.1ex}
    \vfill
    \begin{subfigure}{\linewidth}
        \centering
        \includegraphics[width=0.95\linewidth]{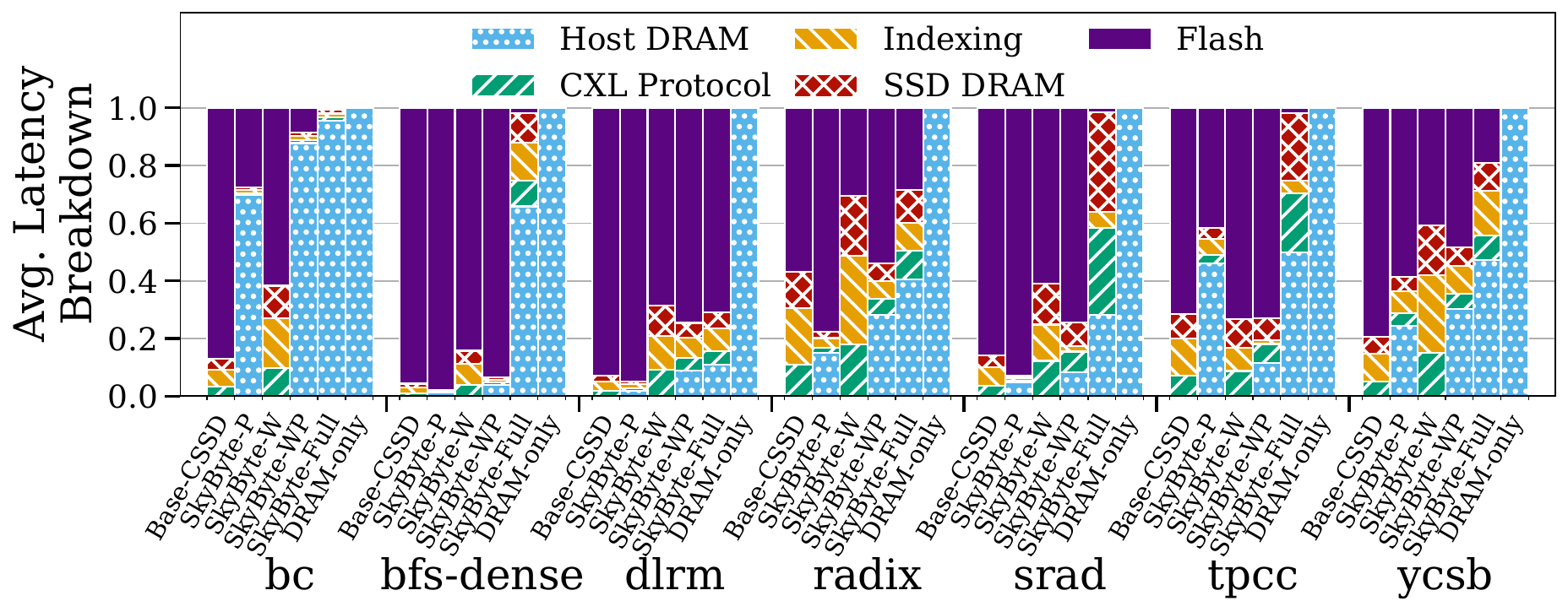}
        \caption{Average memory access time breakdown.}
    \end{subfigure}
    \caption{Average memory access time (AMAT) in \mbox{\pname{}}. (a) compares the AMAT across different designs. (b) shows the percentage breakdown of different AMAT components.}
    \label{fig:eval_amat_breakdown}
\end{figure}


\subsection{Benefit of CXL-Aware SSD DRAM Management}
\label{sec:eval_ssd_cache}
\noindent
\textbf{Average memory access time (AMAT).}
We break down the AMAT of \basename{} and \pname{} variants in Figure~\ref{fig:eval_amat_breakdown}.
To analyze the end-to-end memory latency observed by the host CPU,
we consider the accesses to the promoted pages in the host DRAM as ``host DRAM hits'', while all other accesses go to the CXL-SSD and will suffer the CXL protocol latency (40 ns, see Table \mbox{\ref{tab:exp_setup}}).
For all CXL-SSD accesses, we classify them as ``SSD DRAM hits'' (hits in write log or data cache) or ``SSD DRAM misses'' (which suffer the flash access latency).
The latency of a CXL-SSD access consists of (1) the SSD DRAM cache indexing time (72 ns for the write log, 49 ns for the data cache, see \mbox{$\S$\ref{sec:implementation}}), (2) the SSD DRAM access time (based on the timing in Table \mbox{\ref{tab:exp_setup}}), and (3) the flash chip latency (see Table \mbox{\ref{tab:exp_setup}}) in case of an SSD DRAM miss.
Together, we calculate AMAT by modeling host DRAM, SSD DRAM, and flash chips as a three-level memory hierarchy, where access to SSD DRAM will bypass host DRAM.
For \pname{}-Full, a memory access triggering a context switch is excluded from calculating AMAT since this instruction is squashed. The replayed instruction that eventually retires is included in AMAT.

\begin{figure}[t]
    \centering
    \includegraphics[width=0.93\linewidth]{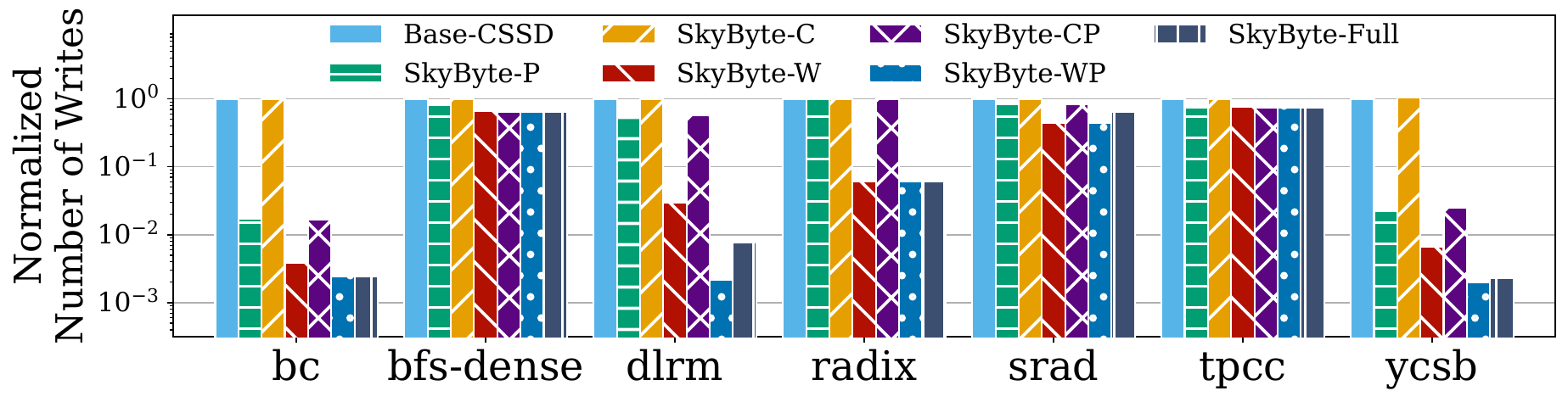}
	\caption{Write traffic of \mbox{\pname{}} to flash chips.}
    \label{fig:eval_write_traffic}
    \vspace{-2ex}
\end{figure}

\begin{figure}[t]
    \centering
    \includegraphics[width=0.93\linewidth]{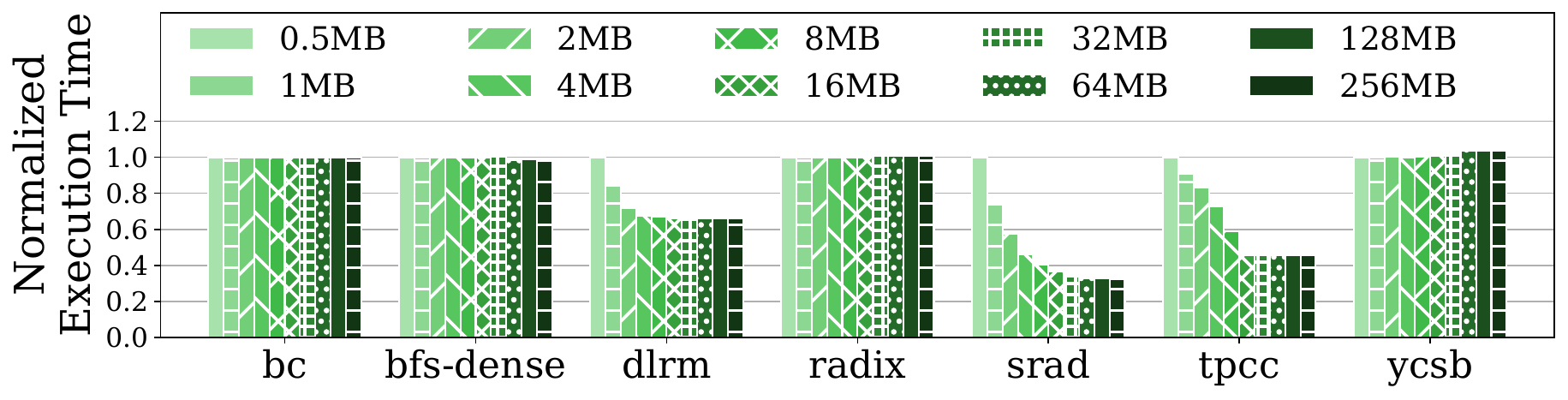}
	\caption{\pname{} performance with various write log sizes.}
    \label{fig:eval_vary_write_log_size}
\end{figure}

\begin{figure}[!tph]
    \centering
    \includegraphics[width=0.93\linewidth]{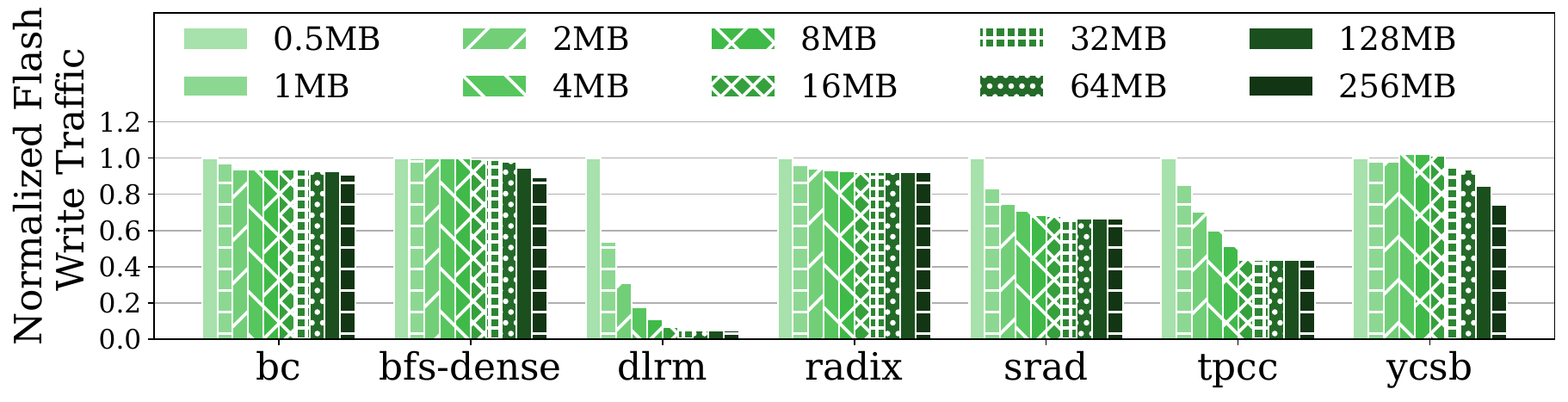}
	\caption{Flash write traffic in \mbox{\pname{}} with various write log sizes.}
    \label{fig:eval_vary_write_log_size_traffic}
      \vspace{-1ex}
\end{figure}

With \pname{}-W, the write log greatly reduces the flash access latency for three reasons. First, writes are buffered in SSD DRAM without expensive flash accesses on the critical path. Second, by coalescing writes to the flash, \pname{}-W reduces the number of flash writes, which leads to less interference to flash reads and triggers GC less frequently. Third, by buffering the writes at cacheline granularity, \pname{}-W caches much fewer unused cachelines.
This improves the efficiency of the SSD DRAM cache.
With \pname{}-P, the page migration reduces the number of SSD accesses, and more memory requests are served by the faster host DRAM.
With both optimizations, \pname{}-WP improves the overall AMAT.


With coordinated context switch, \pname{}-Full achieves better AMAT than \pname{}-WP, because the flash latency can be hidden from applications.
Although the AMAT of \pname{}-Full is still 1.39$\times$ that of the ideal DRAM-Only case on average, the end-to-end performance degradation is only 1.33$\times$ on average (see Figure~\ref{fig:eval_e2e_perf}).
This is because \pname{}-Full exploits the parallelism from more threads than DRAM-Only (24 vs. 8 threads on 8 CPU cores) to tolerate higher AMAT.

\begin{figure*}[t]
    \centering
    \includegraphics[width=0.95\linewidth]{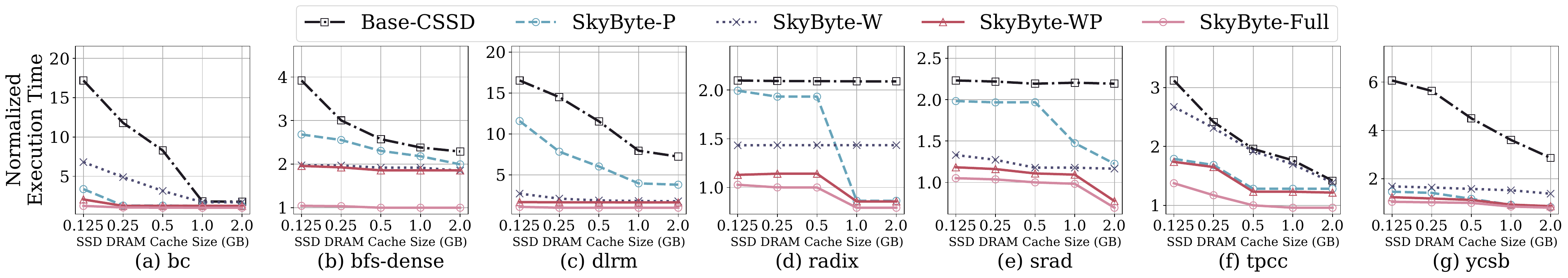}
     \vspace{-1ex}
    \caption{Performance of \mbox{\pname{}} with varying SSD DRAM cache size (normalized to \mbox{\pname{}-Full} with default SSD DRAM cache size). }
    \label{fig:eval_vary_ssd_dram}
    \vspace{-1.5ex}
\end{figure*}

\begin{figure*}[t]
    \centering
    \includegraphics[width=0.95\linewidth]{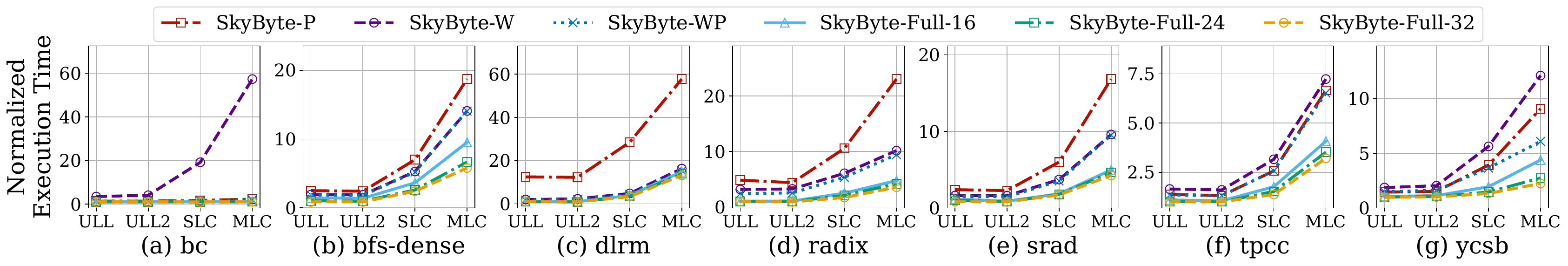}
     \vspace{-1ex}
	\caption{\mbox{\pname{}} performance with varying flash latency. We vary the thread count in \mbox{\pname{}-Full} (e.g. \mbox{\pname{}-Full-16} uses 16 threads).}
    \label{fig:eval_vary_ssd_lat}
      \vspace{-2.5ex}
\end{figure*}


\noindent
\textbf{Flash write traffic.}
Figure~\ref{fig:eval_write_traffic} shows the flash write traffic.
Compared to \basename{}, both \pname{}-P and \pname{}-W reduce the flash write traffic by coalescing writes in the host or SSD DRAM. However, \pname{}-W is more effective because it enables a larger coalescing window with the cacheline-granular write log.
The write log utilizes the limited DRAM space more efficiently, so it can buffer more writes in the limited SSD DRAM.
\pname{}-WP achieves the benefits of both designs.

The context switch mechanism may slightly increase the flash write traffic (e.g., \mbox{\pname{}-Full} v.s. \mbox{\pname{}-WP}, and \mbox{\pname-CP} v.s. \mbox{\pname-P}). This is because context switches enable multiple threads to access the SSD simultaneously, increasing the contention in the SSD DRAM. As a result, the log compaction is triggered more frequently.
This overhead is acceptable given its performance benefits.

\subsection{Impact of Varying Write Log Size}
\label{sec:eval_write_log}

We study the performance impact of varying write log size in Figure~\ref{fig:eval_vary_write_log_size}. We keep the total size of the write log and data cache in the SSD DRAM fixed (i.e., 512MB).
The impact of the write log size mainly depends on the flash write traffic that can be reduced during log compaction, as shown in \mbox{Figure~\ref{fig:eval_vary_write_log_size_traffic}}.
Workloads with more CXL-SSD writes (see Figure~\ref{fig:breakdown_access}) and better temporal write locality will be more sensitive to write log size,  
as a larger write log can coalesce more flash writes (e.g., \texttt{srad} and \texttt{tpcc}).
For most workloads, a small write log (e.g., no more than 64 MB, or 1/8 of the total SSD DRAM size) already provides a sufficiently large write coalescing window.

\subsection{Impact of Varying SSD DRAM Size}
\label{sec:eval_ssd_dram_size}

To further understand the efficiency of the SSD DRAM management in \pname{}, we vary the SSD DRAM data cache size.
In practice, the host DRAM size should scale with a larger SSD DRAM. Hence, we keep the ratio between the maximum size of the promoted pages in the host DRAM and the SSD DRAM cache size the same as the default (i.e., 4:1) in Table~\ref{tab:exp_setup}.
We also keep the ratio between the size of the write log and the data cache to 1:7 for a fair comparison.

Figure~\ref{fig:eval_vary_ssd_dram} shows the performance of \pname{} variants with various SSD DRAM cache sizes.
In all cases, \pname{}-Full is better than all other baselines. The major benefit comes from the efficiency of the cacheline-granular write log, as it enables a larger effective cache size compared to a page-granular cache.
\pname{} helps reduce the cost of CXL-SSDs, as it can achieve similar or better performance with a smaller SSD DRAM compared to \basename{} with a much larger SSD DRAM.




\begin{table}[!t]
\scriptsize
\centering
\caption{The NAND flash parameters used in our evaluation.}
\begin{tabular}{|c|c|c|c|c|}
\hline
\begin{tabular}[c]{@{}c@{}}NAND\\ Type\end{tabular} &
  \begin{tabular}[c]{@{}c@{}}SSD Device\end{tabular} &
  \begin{tabular}[c]{@{}c@{}}Read \\ Time\end{tabular} &
  \begin{tabular}[c]{@{}c@{}}Program\\ Time\end{tabular} &
  \begin{tabular}[c]{@{}c@{}}Erase\\ Time\end{tabular} \\ \hline
ULL  & \begin{tabular}[c]{@{}c@{}}Samsung Z-NAND SSD\cite{znand}\end{tabular} & 3 $\mu$s  & 100 $\mu$s & 1000 $\mu$s \\ \hline
ULL2 & \begin{tabular}[c]{@{}c@{}}Toshiba XL-Flash\cite{toshiba_flash}\end{tabular}   & 4 $\mu$s  & 75 $\mu$s & 850 $\mu$s  \\ \hline
SLC  &                        \cite{atc-cxlssd}                                       & 25 $\mu$s & 200 $\mu$s & 1500 $\mu$s \\ \hline
MLC  &                        \cite{atc-cxlssd}                                       & 50 $\mu$s & 600 $\mu$s & 3000 $\mu$s \\ \hline
\end{tabular}
\vspace{-1ex}
\label{tab:flash_sensitivity}
\end{table}

\subsection{Impact of Varying Flash Latency}
\label{sec:eval_ssd_lat}


Figure~\ref{fig:eval_vary_ssd_lat} shows \pname{}'s performance with different flash chips, including fast ULL flash chips and slower SLC/MLC chips\mbox{~\cite{atc-cxlssd}} (see \mbox{Table~\ref{tab:flash_sensitivity}}).
With a higher flash latency, both \pname{}-WP and \pname{}-Full achieve higher improvement over \pname{}-P. As the benefits of write log and context switching come from the ability to hide flash latency, their benefits are more obvious with a higher flash latency.
\pname{}-Full can utilize more threads to hide the flash latency and improve the application performance.
As long as the SSD bandwidth is enough, \pname{}-Full can scale the performance using more threads until the context switch latency dominates the execution, which aligns with our findings in $\S$\ref{sec:eval_context_switch}.
\pname{} demonstrates that it is promising to use slower yet cheaper commodity flash chips to build CXL-SSDs for parallelizable applications, as they can achieve similar performance with \pname{} compared to the cutting-edge Z-NAND flash chips.



\subsection{Comparison with Alternative Page Migration Mechanisms}
\label{sec:eval_comparison}

Prior studies have proposed various page management mechanisms for tiered memory systems. 
While \mbox{\pname{}} employs an adaptive page migration mechanism (see \mbox{$\S$\ref{subsec:migration}}) by default, in this section, we study the effect of applying alternative page management policies in \mbox{\pname{}}.
We introduce the following designs: (1) \textbf{\mbox{\pname{}}-CT} and (2) \textbf{\mbox{\pname{}}-WCT}. They replace \mbox{\pname{}}'s adaptive page migration with the software-based page migration mechanism of TPP\mbox{\cite{tpp:asplos2023}}.
TPP extends Linux's NUMA balancing and uses periodic sampling and Linux's LRU list to identify hot pages to promote from CXL memory to the host DRAM.
\mbox{\pname{}}-WCT enables the write log in SSD DRAM, while \mbox{\pname{}}-CT does not.
(3) \textbf{AstriFlash-CXL} applies AstriFlash\mbox{\cite{AstriFlash}} on \mbox{\basename{}}.
It employs the host DRAM as a hardware-managed set-associative cache of SSD at 4KB page granularity. It also uses user-level thread switches triggered by host DRAM misses to hide the SSD I/O latency.
Since AstriFlash always accesses the SSD at page granularity, we do not integrate it with \mbox{\pname{}}'s write log design.



\begin{figure}[t]
    \centering
    \vspace{-.5ex}
    \includegraphics[width=0.95\linewidth]{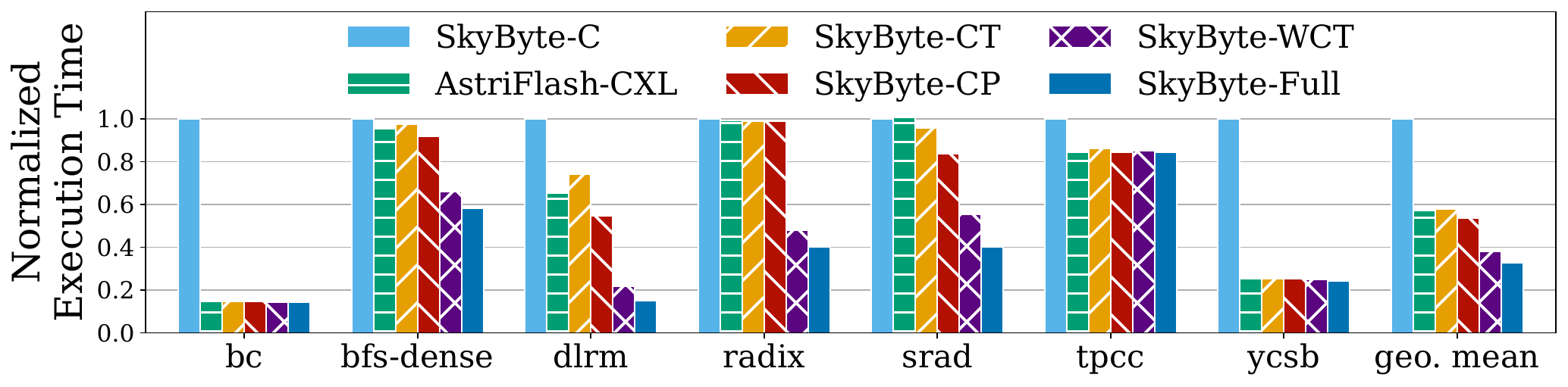}
    \caption{Execution time of \mbox{\pname} with different page migration mechanisms (normalized to \mbox{\pname{}-C}).}
    \label{fig:comparision_tiered}
     \vspace{-1ex}
\end{figure}

Compared to \mbox{\pname{}-C}, all of \mbox{\pname{}-CT}, \mbox{\pname{}}-CP, and \mbox{AstriFlash-CXL} improve performance by caching hot pages in the host DRAM (see Figure \mbox{\ref{fig:comparision_tiered}}). 
\mbox{\pname{}}-CT performs slightly worse than \mbox{\pname{}-CP} on average because TPP uses periodic sampling to estimate page hotness, which is less accurate than the per-page tracking in \mbox{\pname{}}.
\mbox{\pname{}-CP} outperforms AstriFlash-CXL by up to 1.21$\times$ (1.09$\times$ on average), as \mbox{\pname{}} utilizes the host DRAM more efficiently since it only promotes hot pages and effectively uses the host DRAM as a fully associative cache. In contrast, AstriFlash relies on on-demand paging and manages the host DRAM as a set-associative cache.
\mbox{\pname-WCT} outperforms \mbox{\pname{}}-CT by 1.10$\times$ on average, demonstrating that \mbox{\pname{}}'s CXL-aware SSD DRAM management design can be applied to improve the performance of TPP with CXL-SSD. \mbox{\pname{}}-Full further improves over \mbox{\pname{}-WCT} by 1.16$\times$ on average.

\section{Related Work}
\label{sec:related}

\noindent
\textbf{CXL-based Memory Architecture.} 
The recent development of CXL technology~\cite{cxl-spec} 
motivates researchers to rethink the design and deployment of existing memory techniques~\cite{tpp:asplos2023, disaggregated-mem, atc-cxlssd, dsm-sosp2023, pond:asplos2023, cxl-study:micro2023, cxl-anns:atc2023}. For instance, they investigated the impact of CXL on disaggregated memory~\cite{disaggregated-mem, pond:asplos2023, directcxl:atc2022}, and identified new opportunities to facilitate the deployment of tiered memory~\cite{tpp:asplos2023, intel-flat-mode}. CXL has also been applied to SSDs~\cite{atc-cxlssd}. We target such a new architecture, investigate its bottlenecks, and propose a holistic approach to make CXL-based SSDs truly usable with software/hardware co-design.

\noindent
\textbf{Memory-Semantic SSDs.}
Using SSDs to expand main memory capacity has been explored in prior studies~\cite{FlashVM, MMIOSSD,FlashMap, gordon}.
However, they still treated SSDs as block devices and relied on OS paging mechanisms to manage the data movement between the SSD and host memory. A few studies exploited the byte-accessibility of SSDs~\cite{2BSSD, flatflash:asplos2019} and examined its impact on software systems such as ByteFS~\cite{bytefs:asplos2025}. 
\pname{} advances the CXL-SSD with new OS and storage architecture supports. 

\noindent
\textbf{New and Emerging Memory Architecture.}
To overcome the memory wall, alternative memory technologies such as non-volatile memories have been developed~\cite{nvlogging, hoop:isca2020, uniheap:systor2021, ddp:micro2021}. 
Compared to these memory technologies, memory-semantic SSDs provide the generic memory interface while offering scalable memory capacity with much lower cost. To enhance SSD performance, prior studies have developed various architecture-level~\cite{FlashMap, leaftl:asplos2023} and device-level optimization techniques~\cite{znand, somlread:asplos2019}.
\pname{} is compatible with these works, with a focus on addressing the performance challenges in the context of CXL.
Prior studies have proposed informing memory operations that enable the software to proactively react to specific memory events such as cache misses\mbox{\cite{inform_mem_op:isca96}}. They provided new instruction primitives and hardware support for informing memory events. Inspired by these studies, \mbox{\pname{}} uses the SSD controller to inform the host CPU when a long flash access will happen.
AstriFlash\mbox{\cite{AstriFlash}} uses the host DRAM as a hardware-managed cache of the SSD, and hides the flash latency using user-level thread switches triggered by host DRAM misses. However, it still treats the SSD as a black box and manages it at page granularity. \mbox{\pname{}} co-designs the host and the SSD to address the challenges of CXL-SSDs.



\section{Conclusion}
\label{sec:conclusion}
We investigate the performance bottlenecks of CXL-based SSDs and their impact on the application performance. 
We employ a holistic approach to develop \pname{} by co-designing the host OS and SSD, with coordinated context switch and CXL-aware SSD DRAM management. 
We show that \pname{} outperforms current CXL-based SSDs by 6.11$\times$ on average.

\section*{Acknowledgements}
We thank the anonymous reviewers for their insightful comments and feedback. We thank Yuan Xu for his help with memory trace collection. We thank Alaric Yuxiang Chen for his help with our CXL-SSD simulator and discussions at the early stage of this project. This work was partially supported by NSF grant CCF-2107470.


\balance
\bibliographystyle{IEEEtranS}
\bibliography{references}

\newpage
%
%
%
%
%

\appendix

\lstdefinestyle{BashStyle} {frame=tb,
  language=bash,
  aboveskip=3mm,
  belowskip=3mm,
  showstringspaces=false,
  columns=flexible,
  basicstyle={\scriptsize\ttfamily},
  numbers=left,
  numbersep=5pt,
  numberstyle=\tiny\color{gray},
  keywordstyle=\color{blue},
  commentstyle=\color{OliveGreen},
  stringstyle=\color{violet},
  breaklines=true,
  breakatwhitespace=true,
  tabsize=4,
  classoffset=0,
  morekeywords={apt, apt-get, pip3, make, cd, wget},
  keywordstyle=\color{blue},
  classoffset=1,
  morekeywords={optional},
  keywordstyle=\color{orange},
  classoffset=2,
  morekeywords={python3, \text{artifact\_run.sh}}, 
  keywordstyle=\color{violet},
}

\section{Artifact Appendix}

\subsection{Abstract}

 We implemented \pname with a cycle-accurate simulator based on MacSim and SimpleSSD. This artifact includes the source code for \pname's simulation framework and detailed instructions for reproducing the key performance results presented in our paper.
 
 This artifact can run on any x86 machine with at least 32 GB of RAM and 128 GB of disk space. For optimal performance, we recommend using a workstation with multiple high-performance CPU cores and at least 64 GB of RAM. The artifact requires a Linux environment (preferably Ubuntu 20.04 or later) and a compiler supporting the C++11 standard.

\subsection{Artifact check-list (meta-information)}

{\small
\begin{itemize}
  \item {\bf Algorithm:} The threshold-based context-switch trigger policy
  \item {\bf Program:} Benchmarks from Rodinia, GAP, Splashv3, WHISPER, and DLRM. Their traces are included in the artifact.
  \item {\bf Compilation:} g++ 11.4.0 or newer versions.
  \item {\bf Model:} The meta DLRM model. Its traces are included.
  \item {\bf Run-time environment:} Ubuntu 20.04 or newer versions. 
  \item {\bf Metrics:} Execution time, flash write traffic, and average memory access time.
  \item {\bf Output:} Files and graphs, expected results included.
  \item {\bf Experiments:} Generate experiments using provided scripts.
  \item {\bf How much disk space required (approximately)?:} 128 GB.
  \item {\bf How much time is needed to prepare workflow (approximately)?: } Around 40 minutes.
  \item {\bf How much time is needed to complete experiments (approximately)?:} 3 days on a server with 32 CPU cores.
  \item {\bf Publicly available?:} Yes.
  \item {\bf Code licenses (if publicly available)?:} Apache-2.0.
  \item {\bf Data licenses (if publicly available)?:} Apache-2.0.
  \item {\bf Archived (provide DOI)?:} 10.5281/zenodo.14660184.
\end{itemize}
}

\subsection{Description}

\subsubsection{How to access}
The source code can be downloaded from Zenodo at \url{https://zenodo.org/records/14660185}. For the latest version, you can access our Github repo: \url{git@github.com:platformxlab/skybyte.git}.

\subsubsection{Hardware dependencies}

This artifact can run on any x86 machine with a minimum of 32 GB of RAM and 128 GB of disk space.

\subsubsection{Software dependencies}

The artifact requires a Linux environment (preferably Ubuntu 20.04 or later) and a compiler that supports the C++11 standard.

\subsection{Installation}

\begin{enumerate}
    \item Start by downloading the \pname{} artifact from Zenodo:
    \begin{lstlisting}[style=BashStyle,escapechar=~~,label=code:instrumentation]
    wget https://zenodo.org/records/ 14660185/files/SkyByte-Artifact.tar.gz
    tar -xvf SkyByte-Artifact.tar.gz
    \end{lstlisting}
    \item Please make sure all prerequisites are successfully installed:
    \begin{lstlisting}[style=BashStyle,escapechar=~~,label=code:instrumentation]
    sudo apt update
    sudo apt-get install libboost-all-dev
    sudo apt install scons htop
    sudo apt upgrade g++
    pip3 install matplotlib networkx pandas PyPDF2 gdown scipy
    \end{lstlisting}
    \item Build the simulator for \pname{}:
    \begin{lstlisting}[style=BashStyle,escapechar=~~,label=code:instrumentation]
    cd SkyByte-Artifact
    python3 build.py macsim.config -j NUM_THREADS
    # e.g., python3 build.py macsim.config -j 30
    \end{lstlisting}
\end{enumerate}

\subsection{Experiment workflow}

This section describes the steps required to generate and execute the necessary experiments. We strongly recommend referring to \textit{scripts-skybyte/README.md} for detailed explanations of each script used in this process.

\noindent \textbf{Preparing the Multi-threaded Instruction Traces.}

We provide the instruction traces captured using Intel's PIN tool for the workloads discussed in the paper. The traces can be downloaded from Zenodo:

\begin{lstlisting}[style=BashStyle,escapechar=~~,label=code:launch_experiment]
    wget https://zenodo.org/records/ 14660185/files/skybyte_new_traces.tar.gz
    tar -xvf skybyte_new_traces.tar.gz 
\end{lstlisting}

After extracting the files, ensure that the \textit{skybyte\_new\_traces} folder and the codebase (the \textit{SkyByte-Artifact} folder) are located in the same directory.

Each set of traces (e.g., the traces for \texttt{bc} with 16 threads) includes a trace configuration file (\textit{trace.txt}) and several raw trace files (\textit{trace\_XX.raw}). The trace file format is consistent with that of the Macsim simulator. For more details, refer to Section 3.4 of \textit{doc/macsim.pdf}.

\noindent \textbf{Configuration Files.}

The \textit{configs} directory contains configuration files tailored for various workloads, design baselines, and specific settings (e.g., context-switch policies). For detailed information about these files, refer to \textit{configs/README.md}.

\noindent \textbf{Launching A Single Experiment.}

After compiling the simulation framework, a symbolic link named \textit{macsim} will appear in the \textit{bin} directory. Within the same directory, the file \textit{trace\_file\_list} specifies the location of the instruction trace configuration file (i.e., the corresponding \textit{trace.txt}). This artifact includes scripts to automate the setup of individual experiments, which are described in a later section.

To launch a single experiment, use the following command:

\begin{lstlisting}[style=BashStyle,escapechar=~~,label=code:launch_experiment]
    cd bin
    ./macsim -b ../configs/baselines/XX.config -w ../configs/workloads/XX.config (-t ../configs/settings/XX.config) -c {corenum} -o {terminal} -p -f {outputfile_name} (-d) (-r)
\end{lstlisting}

The command-line arguments are defined as follows:
\begin{lstlisting}[style=BashStyle,escapechar=~~,label=code:command_args]
    -b baseline_setting_config_file_name
    -w workload_config_filename
    -t additional_setting_config_file_name (optional)
    -c number_of_logical_cores_to_simlute
    -o terminal_for_printing_warmup_logs (e.g. /dev/pts/6)
    -p: print detailed runtime information (optional)
    -f output_file_name
    -d: run with infinite host DRAM (optional)
    -r: output DRAM-only performance results (optional)
\end{lstlisting}

This command sets up the specified configurations (e.g., design baseline), performs a warmup, and replays the instruction traces on multiple simulated CPU cores and the simulated CXL-SSD. Results will be stored in the \textit{output} directory.

\noindent \textbf{Launching Batched Experiments.}

To execute a large number of experiments simultaneously, we provide the \textit{scripts-skybyte/run\_all.sh} shell script. This script uses regular expressions to match multiple configuration files and automatically spawns experiments in separate \texttt{tmux} windows for parallel execution.

For convenience, we also provide the \textit{artifact\_run.sh} script, which automates the setup and execution of all required experiments. To launch all experiments, simply run:

\begin{lstlisting}[style=BashStyle,escapechar=~~,label=code:run_artifact]
    ./artifact_run.sh
\end{lstlisting}

The variable \texttt{MAX\_CORES\_NUM} in the script specifies the maximum number of CPU cores allowed for simulations. Users may need to adjust this value based on their machine's specifications before running the script.

The \textit{artifact\_run.sh} script performs the following tasks:
1. Creates multiple directories named \texttt{bin-<workload>-<thread\_num>-<baseline>} for different experiments.
2. Sets up the corresponding \textit{trace\_file\_list} file in each directory.
3. Generates a \textit{run\_one.sh} script in each directory to facilitate running individual experiments.
4. Uses the \textit{run\_all.sh} script to launch parallel experiments.

See lines 23-29 of the \textit{artifact\_run.sh} script:

\begin{lstlisting}[style=BashStyle,escapechar=~~,label=code:run_experiment]
    # Set up experiment configurations for figures 2, 3, 4, 14, 15, 16, 17, 18, and Table 3
    ./run_full.sh
    
    # After running this, a folder named bin-<workload_name>-<thread_num>-<baseline_name> will be created for each experiment
    # Inside each folder, there will be a script named run_one.sh to run the individual experiment

    # Run experiments for figures 2, 3, 4, 14, 15, 16, 17, 18, and Table 3 concurrently using multiple cores
    ./run_all.sh -p "bc|tpcc|srad|radix|ycsb|dlrm|bfs-dense" -dr -j $MAX_CORES_NUM
\end{lstlisting}

\subsection{Evaluation and expected results}

To evaluate the artifact results, run the following command:

\begin{lstlisting}[style=BashStyle,escapechar=~~,label=code:evaluate_results]
    ./artifact_draw_figs.sh
\end{lstlisting}

This script collects all results from the \textit{output} folder and generates the required figures sequentially. A detailed description of each command and the locations of the generated figures is provided within the script.

We provide the expected result data files and figures in the same directory where the figures will be generated. To verify the results, you can compare the generated figures directly with those presented in the paper, or compare the data for each figure with the example results we have provided.

\subsection{Experiment customization}

\noindent \textbf{Custom Simulation Configurations.}

In addition to the provided configurations, users can customize their own configuration files and evaluate them. Below is a list of configurable knobs that can be used to customize experiments:

\begin{enumerate}
    \item \texttt{promotion\_enable}: Enables or disables the adaptive page migration mechanism.
    \item \texttt{write\_log\_enable}: Enables or disables CXL-Aware SSD DRAM management.
    \item \texttt{device\_triggered\_ctx\_swt}: Enables or disables the coordinated context switch mechanism.
    \item \texttt{cs\_threshold}: Defines the threshold for the context switch trigger policy. (Unit: ns)
    \item \texttt{ssd\_cache\_size\_byte}: Specifies the size of the SSD DRAM cache. (Unit: Byte)
    \item \texttt{ssd\_cache\_way}: Defines the associativity of the SSD DRAM cache.
    \item \texttt{host\_dram\_size\_byte}: Specifies the size of the host main memory. (Unit: Byte)
    \item \texttt{t\_policy}: Defines the thread scheduling policy. Options include "RR", "RANDOM", and "FAIRNESS" (CFS).
\end{enumerate}

\noindent \textbf{Capturing Custom Program's Traces.}

Users can generate custom traces for their own programs on their machines. To assist with this, we include a sub-repository called \textit{macsim-x86trace} in the artifact. This sub-repo contains the Intel PIN 3.13 tool and scripts that generate both instruction traces and memory warmup traces, which are required by our simulation framework for a custom application. For detailed instructions on how to generate these traces, refer to \textit{macsim-x86trace/README.md}. Please note that PIN 3.13 only runs on Ubuntu 18.04. Users may need to build a new OS environment for collecting custom traces.


\subsection{Methodology}

Submission, reviewing and badging methodology:

\begin{itemize}
  \item \url{https://www.acm.org/publications/policies/artifact-review-and-badging-current}
  \item \url{https://cTuning.org/ae}
\end{itemize}

\end{document}